\documentclass[acmsmall]{acmart}
\setcopyright{none}
\renewcommand\footnotetextcopyrightpermission[1]{}
\usepackage[utf8]{inputenc}
\usepackage{setspace}
\usepackage{graphicx}
\usepackage{mathtools}
\usepackage{comment}
\usepackage{algorithm,algpseudocode}
\usepackage{subfigure}

\usepackage{multirow}
\usepackage{float}
\usepackage{soul}

\usepackage{amsfonts}
\usepackage{booktabs}
\usepackage{amsmath}
\usepackage{bm}

\usepackage{tabularx}
\usepackage{array}
\usepackage{breqn}
\usepackage{supertabular}

\usepackage{amsthm}
\usepackage{xcolor}
\usepackage{makecell}
\usepackage{ragged2e}  
\newcolumntype{Y}{>{\RaggedRight\arraybackslash}X} 
\theoremstyle{acmplain}
\newtheorem{theorem}{Theorem}[section]
\newtheorem{lemma}{Lemma}[section]
\newtheorem{definition}{Definition}[section]

\newcommand{\R}{\mathbb{R}}

\usepackage{enumitem}

\usepackage[
singlelinecheck=false 
]{caption}
\usepackage[finalnew]{trackchanges}

\begin{document}

\title{PRECISE: Private Regulatory Compliance for Cyberattack Detection on Critical Infrastructure Systems}





\author{Sathwik Yamana}
\email{syamana@okstate.edu}

\author{Paritosh Ramanan}
\email{paritosh.ramanan@okstate.edu}

\author{H.\,M. Mohaimanul Islam}
\email{h\_m\_mohaimanul.islam@okstate.edu}

\author{Abhiram Reddy Alugula}
\email{aalugul@okstate.edu}

\affiliation{
  \institution{Oklahoma State University}
  \department{School of Industrial Engineering and Management}
  \city{Stillwater}
  \state{Oklahoma}
  \country{USA}
}
\thanks{Under review at ACM Transactions on Privacy and Security.}

\begin{abstract} 
Industrial control systems are a fundamental component of critical infrastructure networks (CIN) such as gas, water and power. With the growing risk of cyberattacks, regulatory compliance requirements are also increasing for large scale critical infrastructure systems comprising multiple utility stakeholders. The primary goal of regulators is to ensure overall system stability with recourse to trustworthy stakeholder attack detection. However, adhering to compliance requirements requires stakeholders to also disclose sensor and control data to regulators raising privacy concerns. In this paper, we present a cyberattack detection framework PRECISE, that utilizes differentially private (DP) hypothesis tests geared towards enhancing regulatory confidence while alleviating privacy concerns of CIN stakeholders. The hallmark of our approach is a two phase privacy scheme applying Laplacian DP to covariance matrix disclosures and Gaussian Differential Privacy (GDP) to state-space residuals derived from a Non-Linear Kalman Filter LSTM model. We formally characterize the GDP-induced test statistic via the non-central chi-squared distribution and derive tight bounds on misclassification rates and equivalent DP levels of significance. Theoretically, we show that our method induces a misclassification error rate comparable to the non-DP cases while delivering robust privacy guarantees.Using real-world HAI and ORNL-PS datasets, we demonstrate that under strong differential privacy guarantees on both covariance and residual disclosures, PRECISE matches non-DP detection outcomes in over 88\% of cases within 600 seconds of attack onset for the HAI dataset and over 92\% for the ORNL-PS dataset, while maintaining false alarm rates below 9\% across all tested DP parameter configurations.
\end{abstract}

\maketitle
\section{Introduction}

Large-scale Critical Infrastructure Networks (CINs) are characterized by physically interdependent subsystems of varying network sizes that are operated by a diverse group of utility stakeholders governed by a regulatory entity. Data-driven cyberattacks targeting key operational technology (OT) components, such as industrial control systems (ICS), of several utilities, have been shown to cause devastating cascading failures that threaten overall network stability \cite{bompard2011structural}. In the United States, there have been several attempts aimed at establishing information-sharing and analysis centers (ISACs) as a means for detecting network-wide ICS attacks through secure aggregation of sensor data as well as local cyber-incident alarms \cite{smith2023cybersecurity}. However, efforts to establish regulatory compliance initiatives like ISACs have severely fallen short of expectations primarily due to the tepid response from stakeholders \cite{dhs_oig}. Privacy concerns of utilities as well as lack of trust and credibility of reported information form the core set of obstacles that threaten the feasibility of ISAC-like compliance frameworks. Therefore, in this paper, we introduce PRECISE a differentially private subsystem-level ICS attack detection framework that relies on privacy-preserving disclosures of underlying datasets by utilities. Additionally, PRECISE enables ISAC-like entities to verify compliance of reported detection outcomes with respect to the disclosed datasets, leading to increased trust and credibility in regulatory outcomes.

Enhancing the cyber resilience of interdependent CINs in order to limit disruptions from cascading impacts is a key research priority, as outlined by the Cybersecurity and Infrastructure Security Agency (CISA) \cite{ripdw}. In such cases, regulatory entities like ISACs play a critical role in stemming the impacts of data-driven ICS attacks by coordinating cyber-incident response \cite{dhs_oig} accompanied by timely dissemination of insights. However, the capabilities of ISACs are only as good as the quality of the alarms and the associated datasets reported from the utility stakeholders. In fact, poor quality data collected by ISACs without sufficient context impedes situational awareness, hampers decision-making, and increases false positives occasionally resulting in unnecessary system upgrades \cite{dhs_oig2}. Therefore, enabling ISACs to verify that the reported alarms comply with the underlying datasets of the utilities can help pave the way for improved situational awareness and reduced false alarm rates across the entire network. 

The ability to verify compliance of alarms on the basis of underlying datasets is challenging due to privacy concerns of utility stakeholders. Privacy concerns also impede real-time information sharing among CIN stakeholders \cite{nolan2015cybersecurity}. It has been demonstrated that operational data from CINs can be used to identify industrial customer demands \cite{mak2019privacy}, reveal operational costs of strategic CIN \cite{fioretto2019ppsm}, and identify systemic vulnerabilities \cite{fioretto2019privacy}. There is also considerable trepidation among stakeholders regarding the perceived misuse of information obtained from cyber information-sharing programs \cite{nolan2015cybersecurity} by governmental agencies. In some cases CIN stakeholders are also concerned about risks to their organizational reputation \cite{nweke2020legal} in the event of data leaks. As a result, the need of the hour is a private, trustworthy framework for detecting data-driven ICS attacks \cite{johnson2016guide} while enabling data-driven compliance verification of the reported alarms.

Conventionally, ICS frameworks capture IoT and sensor data from assets \cite{li2020detection} across each utility. ICS frameworks utilize state-space models to characterize the operational aspect of utilities \cite{li2020deep} and detect anomalies and deviations from steady state conditions. A majority of these anomaly detection frameworks utilize statistical hypothesis tests on state space residuals computed using sensor data \cite{li2022online}. As a result, alarms at the utility level are an outcome of anomaly detection frameworks that is inherently characterized by local sensor data as well as the associated covariance matrices. Statistical analysis of residuals can also yield significant insights into the nature of detected anomalies such as attack diagnosis and distinguishing attacks from routine equipment failure \cite{li2022online}. As a result, statistical methods are widely used as a first line of defense in identifying potential attack driven anomalies in ICS frameworks. 

In this paper, we consider the interaction between the utility stakeholders and their regulatory counterparts such as ISACs. The goal of the utility is to \textit{convince} an ISAC regarding alarm validity through privacy-preserving disclosures of the underlying residuals and the detection algorithm used for alarms.
\begin{figure}
\includegraphics[width=0.45\textwidth]{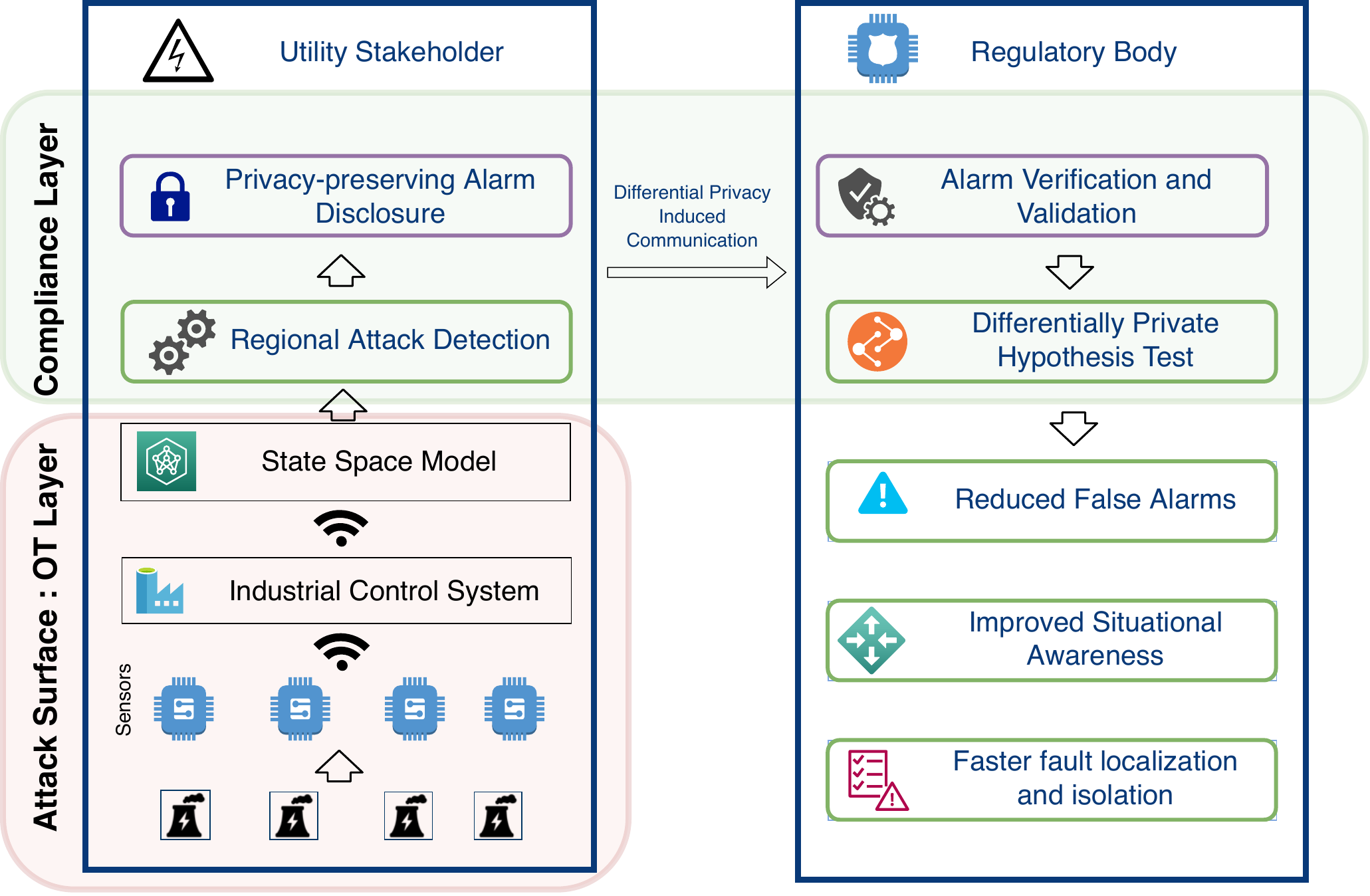}
\caption{\label{fig:threat_model} Compliance-driven, private ICS attack detection}
\vspace{-5mm}
\end{figure} 
Our proposed approach therefore uses differential privacy (DP) driven disclosures of data as a means to drive the transparent validation of the alarms. A schematic of our proposed approach is depicted in Figure \ref{fig:threat_model}. However, two major bottlenecks arise with this approach. First, statistical hypothesis tests, commonly used for data-driven ICS attack detection, require privacy-preserving disclosures of local covariance matrices to generate alarms \cite{li2020detection}. Second, using differentially private data disclosures can itself severely affect the alarm outcomes leading to misclassification with respect to original alarm statistics \cite{couch2019differentially,gaboardi2018local}. Consequently, ensuring compliance and transparency through data sharing introduces a high degree of statistical complexity driven by privacy constraints.

In this paper, we target both these bottlenecks by adopting a two phase DP mechanism to introduce transparency and regulatory compliance for ICS based attack detection. In the first phase, we focus on DP based disclosures of covariance matrices in order to characterize the utility-level detection framework. The second phase involves privacy-preserving disclosures of temporal state-space residual values by leveraging Gaussian differential privacy frameworks. We derive $(\epsilon,\delta)$ differential privacy guarantees on data-driven ICS attack detection frameworks. Our contributions can be summarized as follows:
\begin{itemize}
\item We develop an algorithmic framework for privacy-preserving hypothesis tests that are compatible with DP disclosures of covariance and state-space residuals.
\item We develop two distinct implementation modes that enable regulatory bodies like ISACs to independently verify utility detection outcomes using DP disclosures of high-dimensional ICS data.
\item We derive strong privacy guarantees governing residual and covariance disclosures in order to obtain DP equivalent levels of significance and test statistic distributions for the corresponding hypothesis tests. 
\item We theoretically characterize the impact of DP on detection quality by analyzing the DP-induced levels of significance and associated test statistics.
\end{itemize}
Our proposed framework is evaluated using a generalizable state-space modeling framework that utilizes a non-linear Kalman Filter based approach. Our experimental results are demonstrated using real-world ICS data \cite{hai} consisting of diverse attack scenarios on several heterogenous subsystems. {The key takeaway of our research is that DP based disclosures offer a viable alternative for establishing regulatory compliance standards, help achieve higher degree of situational awareness, trust and credibility in CINs while providing strong privacy guarantees for utility stakeholders.}   

\section{Related Work}
Attacks like DoS, DDoS, and phishing, which specifically target IT systems, can often be effectively detected and isolated by monitoring network traffic \cite{lazarevic2003comparative,caselli2016specification,ye2004robustness}. However, data-driven attacks on ICS are specifically designed to evade such conventional defense mechanisms, including network monitoring and protocol-based intrusion detection \cite{kang2014cyber,drias2015taxonomy}. These attacks form a more significant threat to ICSs due to their ability to impact information, communication, and the underlying physical systems \cite{urbina2016limiting}, leading to significant damage to critical infrastructure. Unlike IT-layer attacks, sensor data is manipulated in order to effect damage through malicious control actions and incorrect state estimations, resulting in degraded asset performance.


Several model-based detection mechanisms have been proposed for ICS attacks involving data manipulation \cite{liu2011false,dan2010stealth, teixeira2010cyber, yu2015blind, cardenas2011attacks,chaojun2015detecting,mo2014detecting, huang2018online,van2015sequential,hoehn2016detection,weerakkody2015detecting,pan2015classification,rahman2017multi}. A popular state-space estimation modeling framework used for ICS attack detection is the Kalman Filter based technique \cite{li2020deep,li2020detection,li2021degradation,ramanan2021blockchain, li2022online}. The Kalman Filter based detection algorithms rely on residuals computed on the basis of the estimated (or predicted) and observed measurements or states followed by a statistical testing procedure \cite{irita2017detection,danLi}. Due to the presence of robust physics-based models in industrial IoT, the Kalman Filter based methods form a powerful class of techniques for anomaly detection. Using a Kalman Filter based state space estimation, one can conduct attack diagnosis \cite{li2022online}, distinguish between routine faults and attacks using degradation models \cite{li2021degradation} and use decentralized methodologies to raise network-wide alarms \cite{ramanan2021blockchain}. However, detecting ICS attacks across multi-stakeholder CINs requires modeling spatiotemporal interdependencies among utilities \cite{palleti2021cascading, bompard2011structural}, introducing significant privacy concerns when operational data must be shared.


On the other hand, recent works have investigated the use of DP as a means to circumvent the privacy-related impediments to enable data sharing in CINs \cite{fioretto2019privacy}. DP is a widely used method to protect the privacy of data sets intended to be communicated through public domains \cite{ dwork2014algorithmic,cortes2016differential}. DP-driven approaches involve injecting a randomized noise in order to obfuscate the real underlying data record \cite{zhang2018enabling, cao2020differentially}. The injected randomized noise can be designed so as to facilitate theoretical guarantees bounding the loss of privacy \cite{dwork2014algorithmic}. DP thereby ensures that the probability of extracting the real value from a noisy data set by any external entity remains remarkably low. Most DP approaches applied in the context of CINs are geared toward the public release of operational data for benchmarking purposes \cite{zhou2019differential,fioretto2018constrained,fioretto2019differential,mak2019privacy,fioretto2019privacy,mak2020privacy,dvorkin2020differentially}. The use of DP for protecting the input signals in a Kalman Filter based state space modeling framework has also received considerable attention as well \cite{le2020differentially,song2018privacy,degue2017differentially}.

Additionally, there have been several approaches that have focused on differential privacy in the context of statistical hypothesis tests \cite{couch2019differentially,gaboardi2016differentially}. Such methods have typically relied on developing DP versions of popular statistical hypothesis tests such as Wilcoxon-signed test \cite{couch2019differentially}, goodness-of-fit tests\cite{gaboardi2016differentially,gaboardi2018local,rogers2017new}, F-test for linear regression significance \cite{alabi2022hypothesis}. However, a majority of them target categorical datasets \cite{gaboardi2018local, rogers2017new} or are nonparametric in nature \cite{couch2019differentially}. Therefore, while most of these methods provide strong methodological foundations, they are not geared towards temporal datasets that result from state-space modeling approaches such as Kalman Filter. \textit{As a result, there exists a critical methodological gap for DP based approaches that can specifically cater to anomaly detection for temporal state-space models.}

Secure multiparty computation (MPC), homomorphic encryption (HE), and federated learning (FL) have all been explored for privacy-preserving data sharing in CINs. However, each comes with practical limitations that make them unsuitable for our setting. MPC and HE fundamentally require coordinated computation between parties, which introduces significant communication overhead that is difficult to sustain in real-time ICS monitoring environments. Privacy-preserving federated learning paradigms, while well-suited for distributed model training, offer no benefit for enabling compliance between the regulators and stakeholders in the context of anomaly detection in CINs. Moreover, federated learning paradigms are unable to provide provable worst-case guarantees on the privacy of individual data when considering regulatory disclosures. Differential privacy, on the other hand, provides mathematically rigorous bounds on privacy loss, operates without requiring coordination between utilities, and is computationally lightweight enough to handle sequential temporal data — making it the most appropriate choice for privacy-preserving disclosures of state-space residuals and covariance matrices in the regulatory compliance setting considered in this paper.


\section{State Space Modeling for ICS}\label{sec:ssm}
Operational modeling of utility stakeholder level ICS is the preliminary step for building a robust detection framework that can be ultimately leveraged for regulatory compliance. In that regard, an exceptional operational model must possess two critical estimation capabilities. First, operational models of utility ICS must incorporate transition functions that can help estimate the future state of the system based on existing sensor data. Second, such models must also be capable of yielding stable and accurate estimations of process and sensor noise distributions which are critical in helping make future state estimations more robust. The core idea is that an exceptional operational model can be utilized to statistically distinguish between normal and anomalous ICS behavior enabling accurate local attack detection which will in turn drive regulatory compliance. For ease of reference, Table~\ref{tab:notation} provides a summary of the key notation used throughout this paper.

\begin{table}[t]
\centering
\small
\caption{Summary of Notation}
\label{tab:notation}
\begin{tabular}{ll}
\toprule
\textbf{Symbol} & \textbf{Definition} \\
\midrule
\multicolumn{2}{l}{\textit{State Space / Kalman Filter}} \\
\midrule
$x_t$ & Latent state vector \\
$y_t$ & Sensor measurements \\
$u_t$ & Control action \\
$v_t, w_t$ & Process and measurement noise \\
$Q_t, R_t$ & Process and measurement noise covariance \\
$S_t$ & Residual covariance matrix \\
$r_t$ & Residual vector \\
$K_t$ & Kalman gain \\
$P_{t|t}, P_{t|t-1}$ & Posteriori and priori state covariance \\
$G_t, H_t$ & State transition and observation Jacobians \\
$\tau_t$ & Whitened residual vector \\
$T_{\chi^2,t}$ & Original test statistic $\|\tau_t\|_2^2$ \\
$\rho_t$ & Original (non-DP) alarm \\
\midrule
\multicolumn{2}{l}{\textit{Differential Privacy Parameters}} \\
\midrule
$\epsilon_{\mathrm{cov}}, \epsilon_r$ & Privacy budgets for covariance and residual \\
$\gamma_{\mathrm{cov}}, \gamma_r$ & DP failure probabilities \\
$\Delta_l, \Delta_r$ & $\ell_1$ and $\ell_2$ sensitivities \\
$\sigma$ & GDP noise scale \\
$\theta_l, \theta_r$ & DP noise bound parameters \\
\midrule
\multicolumn{2}{l}{\textit{Perturbed Quantities}} \\
\midrule
$\hat{S}_t$ & DP perturbed covariance matrix \\
$\hat{\tau}^{\mathrm{cov}}_t$ & Covariance DP perturbed residual \\
$\hat{\tau}^{\mathrm{res}}_t$ & GDP perturbed residual \\
$\hat{T}^{\mathrm{cov}}_{\chi^2,t},\, \hat{T}^{\mathrm{res}}_{\chi^2,t}$ & Perturbed test statistics \\
$\hat{\alpha}$ & DP equivalent level of significance \\
$\hat{\rho}_t$ & DP alarm \\
\midrule
\multicolumn{2}{l}{\textit{Misclassification / Compliance}} \\
\midrule
$R_t$ & Squared residual norm $\sum_{i=1}^p (r^i_t)^2$ \\
$\Pi_t$ & Residuals and DP parameters at time $t$ \\
$\mu_t$ & GDP noise calibration factor \\
$\chi^2_\alpha,\, \chi^{2,NC}_{\hat{\alpha}}$ & DP and non-DP quantile thresholds \\
\midrule
\multicolumn{2}{l}{\textit{Algorithmic / Epoch}} \\
\midrule
$w,\, W$ & Epoch index and steps per epoch \\
$\mathcal{J},\, J$ & Set of utilities and cardinality \\
$r_w,\, S_w$ & Epoch-aggregated residual and covariance \\
\bottomrule
\end{tabular}
\end{table}

\subsection{Non-Linear State Space Formulation}
For instituting a local ICS attack detection framework, we begin by discussing a generalizable state-space modeling framework for characterizing utility level ICS operations that are non-linear in nature. Our generalizable state-space model considers a sensor-driven non-linear system at time $t$, where $x_t\in \R^{m}$ represents the latent space embedding, {$u_t\in \R^{m}$} represents the control action and $y_t\in \R^{d}$ represents noisy sensor measurements from asset sensors.
\begin{gather}
    x_{t+1}= g(x_{t-1},u_{t-1}) + v_t,\label{eq:eqn1a}\\
    y_t=h(x_t)+w_t, \label{eq:eqn1b}
\end{gather}
In Equations \eqref{eq:eqn1a}, \eqref{eq:eqn1b}, {$\bm{g}, \bm{h}$} are the state transition and observation functions respectively. Collectively, {$\bm{g}, \bm{h}$} represent the relationship between the measurements $y_t$ and the state $x_t$. The process and measurement noises at time $t$ are denoted by $v_t\in \R^{m}$,  and $w_t\in \R^{m}$ respectively. The process and measurement noises follow multivariate normal distributions with zero mean implying that $v_t \sim N(0,Q_t), w_t \sim N(0,R_t)$, where $Q_t,R_t$ represent the covariance matrices respectively. Such a type of modeling framework has also been used extensively in prior work \cite{smith2011decoupled, mo2014detecting, teixeira2010cyber}. A Non-Linear Kalman Filter can be used to model the state space of the stakeholder ICS as represented by Equations \eqref{eq:eqn1a} and \eqref{eq:eqn1b}. 


\subsection{Non-linear Kalman Filter Estimation}\label{subsec:nlkfe}
In the non-linear model given in Equations \eqref{eq:eqn1a}, \eqref{eq:eqn1b}, an extended Kalman Filer based model denoted by $\mathcal{K}$ can be formulated using the following equations. 
\begin{gather}
    \hat{x}_{t|t-1}=g(\hat{x}_{t-1|t-1},u_{t-1}) \label{eq:KLQG1},\\
    r_t=y_t-h(\hat{x}_{t|t-1}) \label{eq:KLQG2},\\
    P_{t|t-1} = G_tP_{t-1|t-1}G^T_t + Q_{t-1} \label{eq:KLQG3},\\
    S_t = H_tP_{t|t-1}H^T_t+R_t \label{eq:KLQG4},\\
    K_t = P_{t|t-1}H^{T}_{t}S_t^{-1} \label{eq:KLQG5}\\
    \hat{x}_{t|t}=\hat{x}_{t|t-1}+K_tr_t \label{eq:KLQG6}\\
    P_{t|t} = (I - K_tH_t)P_{t|t-1} \label{eq:KLQG7}
\end{gather}
In Equations \eqref{eq:KLQG3},\eqref{eq:KLQG5}, $P_{t|t-1},P_{t|t}$ represent the predicted and the updated covariance estimates respectively, while $S_t$ represents the residual covariance at $t$. The state transition and the observation matrices at $t$ given by $G_t, H_t$ respectively are computed using the Jacobians of $g,h$ as denoted by the following equations.
\begin{gather}
G_t = \left. \frac{\partial g}{\partial x} \right|_{x_{t-1|t-1},u_t}\text{  and  } H_t = \left. \frac{\partial h}{\partial x} \right|_{x_{t|t-1}}
\end{gather}

\subsection{Learning the Non-Linear Kalman Filter}
The stakeholder level ICS system can be modeled as the Non-Linear Kalman Filter (NLKF) described in Section \ref{subsec:nlkfe}. The NLKF model requires a generalizable framework that is known \emph{a priori} and can characterize the stakeholder level ICS system dynamics including the estimation of process and measurement noise covariance estimation based on current and historic sensor measurements and state estimates. 

However, constructing a generalizable framework that efficiently captures ICS system dynamics is challenging due to the tedious nature of estimating and fitting parametrized probability distributions to process and measurement noise with covariance $Q_t,R_t$ respectively. Without accurate knowledge of these covariance matrices, the Kalman gain cannot be computed precisely resulting in erroneous posteriori state estimates predictions which can compromise the detection quality for data driven attacks.  

Therefore, we utilize a machine learning framework comprising of Long and Short Term Memory (LSTM) based recurrent neural networks complemented by feed-forward layers that attempt to cast the ICS process towards a Non-Linear Kalman Filter based setting. More precisely, the Non-Linear Kalman Filter based LSTM (NLKF) design involves the posteriori state estimates of the prior time step $\hat{x}_{t|t}$ in addition to the observed sensor measurements $y_t$ to \emph{predict} the process and measurement noise covariance $Q_t,R_t$.

\begin{figure}[!htb]
\includegraphics[width=0.45\textwidth]{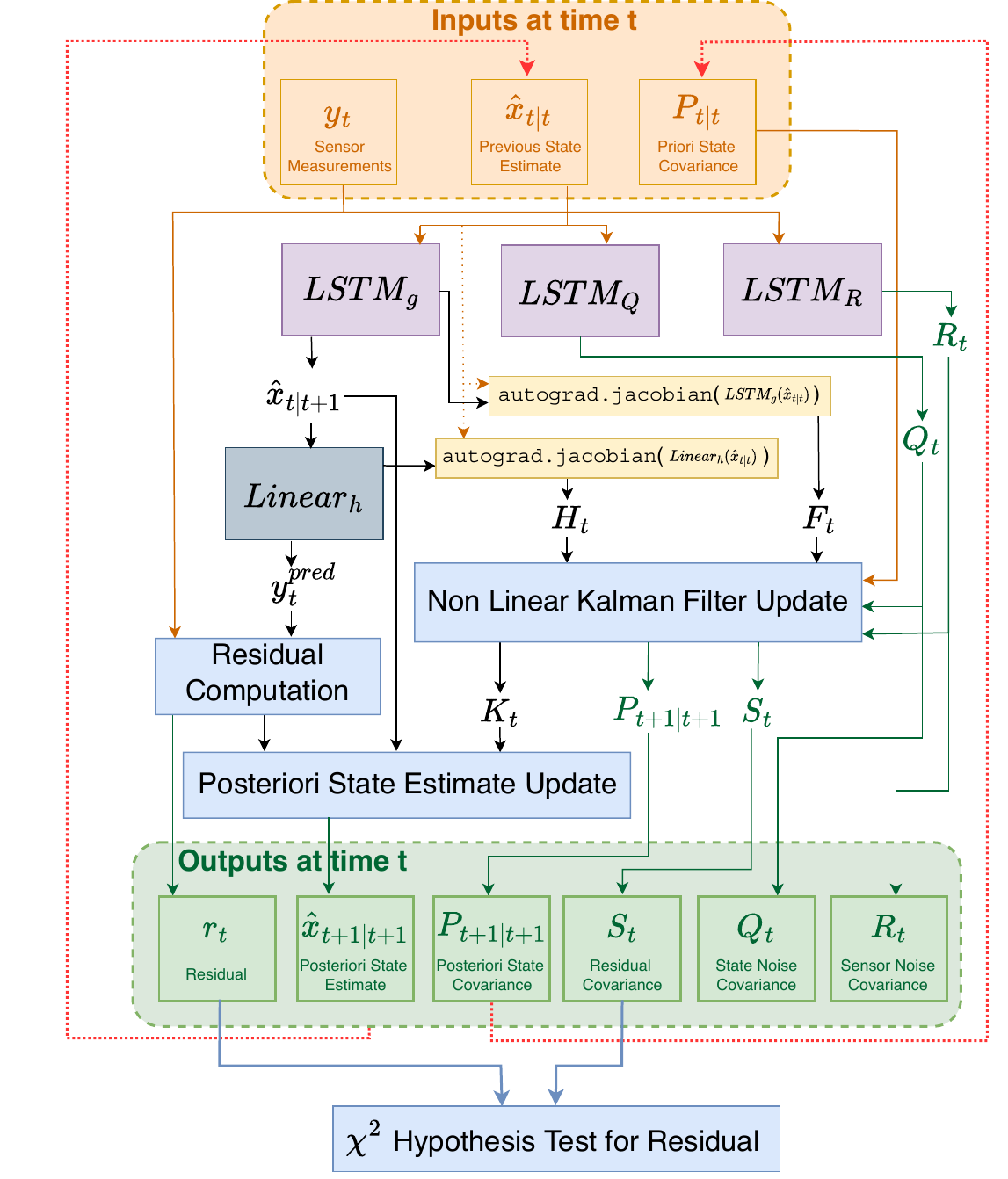}
\caption{\label{fig:nlkflstm} Framework for learning Non-Linear Kalman Filter}
\vspace{-2mm}
\end{figure} 

Figure \ref{fig:nlkflstm} depicts the NLKF framework that is used for learning the state space model. The NLKF framework consists of three LSTM ($LSTM_g$, $LSTM_Q$ and $LSTM_R$) and one feed-forward Linear layers ($Linear_h$) that are collectively used to learn the non-linear interdependencies between the sensor measurements $y_t$, and the priori and posteriori state estimates $\hat{x}_{t|t-1}$ and $\hat{x}_{t|t}$. Specifically, $LSTM_g$ learns the relationship between the current state estimate $\hat{x}_{t|t}$ and the \emph{a priori} state estimate $\hat{x}_{t|t+1}$. $LSTM_Q$ and $LSTM_R$ predict the state and measurement covariance matrices based on $\hat{x}_{t|t}$ and $y_t$ respectively. We use $Linear_h$ to map the lower dimensional \emph{a priori} state estimates to the high dimensional sensor data space in order to predict $y^{pred}_t$ such that residual $r_t = y_t - y^{pred}_t$ can be computed. Using the autograd functionality in machine learning frameworks like PyTorch, we compute the Jacobians $H_t$ and $G_t$ for the $LSTM_g$ and $Linear_h$ respectively. Finally, the estimated Jacobians, sensor noise, state noise and state covariance matrices enable us to calculate the extended Kalman gain matrix $K_t$ and the posterior state covariance $P_{t+1|t+1}$ according to Equations \eqref{eq:KLQG3}-\eqref{eq:KLQG5}. Therefore, the NLKF framework provides us with the fundamental ability to generate residual $r_t$ estimates along with temporally sound predictions of the covariance matrices $Q_t,R_t,S_t$ as well.

\subsection{Hypothesis Tests for Anomaly Detection}\label{subsec:htad}
Estimation of covariance matrices enables us to implement statistical hypothesis tests on the corresponding covariates themselves. As a result, we can leverage the NLKF framework, for orchestrating the $\chi^2$ hypothesis tests, that can detect covert attacks \cite{li2020detection} using $S_t$ and $r_t$. Additionally, the same methodology can also be used for detecting false data injection attacks using sensor data measurements $y_t$ and covariance $R_t$\cite{li2020deep}. 

To establish the theoretical underpinnings of the $\chi^2$ hypothesis test, we consider without any loss of generality, the predicted residual covariance matrix $S_t$ and the residual $r_t$ at time $t$. $S_t$ is symmetric, which means that $S_t = V_t \Lambda_t V_t^{T}$, where $V_t$ describes the set of orthonormal eigenvectors and $\Lambda_t$ is a diagonal matrix representing a set of eigenvalues. As a result, we can obtain the principal component decomposition of $S_t$. We know that $(S_t)^{-1/2} = (V\Lambda_t^{-1}V_t^{T})^{1/2} = V_t\Lambda_t^{-1/2}$. Given a residual vector $r_t\sim N(0,S_t)$, we can realize $\tau_t = (S_t)^{-1/2}(r_t) = S_t\Lambda_t^{-1/2}(r_t)$ such that $\tau_t \sim N(0,I)$ \cite{li2020detection,hwang2017chi}. Therefore, to detect attacks, the utility stakeholder level $\chi^2$-hypothesis test can be formulated as follows:
\begin{gather}
H_0 : S^r_t = S_t \label{eq:nullh}\\
H_A : S^r_t \neq S^{true}_t \label{eq:alth}
\end{gather}
The null hypothesis $H_0$ represents the condition that the covariance of the reported residual vector $r_t$ denoted by $S^r_t$ is equal to the expected covariance $S_t$ predicted by the NLKF framework. In other words, the null hypothesis tests whether the residual vector $r_t$ indeed follows the parametrized distribution $N(0,S_t)$. The underlying insight behind $H_0$ is that perturbations due to abnormal sensor readings or faulty computation of state estimates, will result in residual vectors that adhere to a covariance matrix that is statistically different from the one predicted by the NLKF framework. Further, we can also state that if $H_0$ holds, then the standardized vector of principal component (PC) scores is given by $\tau_t$ implies that $||\tau_t||^2_2 \sim \chi^2_p$, where $p<d$ is the degrees of freedom corresponding to the number of principal components used \cite{hwang2017chi}. 
\begin{equation}
\rho =
\begin{cases}
1, & \text{if } T_{\chi^2,t} = \|\tau_t\|^2_2 > \chi^2_{m,\alpha}, \\
0, & \text{otherwise.}
\end{cases}
\label{eq:alarm_condition}
\end{equation}
Based on Equation \eqref{eq:alarm_condition}, an alarm $\rho=1$ is triggered when the test statistic $T_{\chi^2,t} = ||\tau_t||^2_2 > \chi^2_{m,\alpha}$, where $\alpha$ is the level of significance, while $\rho=0$ otherwise \cite{li2020deep}.

\section{Privacy Preserving Detection Scheme}
Detecting attacks requires continuous monitoring of the residual vector. However, the alarms might need to be publicly validated by a third party (such as a nodal authority or overseer agency) due to regulatory and compliance reasons. As a result, residual vectors and the covariance matrices need to be disclosed to facilitate public validation of alarm values so as to meet regulatory compliance. However, a critical challenge arises in the inability to provide strong guarantees regarding the privacy loss stemming from the public disclosure of these data.

\subsection{Privacy Scope and Objectives}
We focus on a utility based subsystem and its associated state-space model as detailed in Section~\ref{sec:ssm}. Our privacy scope particularly focuses on the disclosure of $S_t$ and whitened residual $\tau_t$ resulting from the NLKF framework as a means to validate the alarm value $\rho_t$. In practice, regulatory compliance for CIN utilities is operationalized through ISAC-driven information sharing frameworks~\cite{smith2023cybersecurity, dhs_oig}, wherein utilities are required to disclose detection outcomes and supporting data to regulatory bodies~\cite{johnson2016guide}. However, privacy concerns~\cite{nolan2015cybersecurity} and lack of trust in disclosed information~\cite{dhs_oig, dhs_oig2} have severely hampered compliance efforts. PRECISE directly addresses these impediments by enabling utilities to satisfy disclosure requirements through differentially private disclosures of $S_t$ and $\tau_t$. As a result, we characterize our threat model and related assumptions as discussed below.

\subsection{Threat Model}
We consider two classes of adversaries relevant to our framework and discuss the relevant trust and security implications of our underlying privacy-preserving detection framework.

\noindent \textbf{Cyber adversary}: The cyber adversary targets the utility-level ICS with the objective of inducing operationally significant damage through manipulation of sensor measurements. Such manipulation can cause the actual residual covariance $S^r_t$ to deviate statistically from the NLKF predicted covariance $S_t$ which will statistically trigger the alternate hypothesis $H_A$. Attack manipulation is assumed to occur only at the sensor level in the ICS hierarchy. We assume that the attacker does not possess visibility into or the capability to manipulate higher levels of the ICS, including the NLKF anomaly detection modules as well as other IT infrastructure itself. We focus on data injection attacks as the modality of the cyber adversary. The treatment of gradual degradation-based attacks \cite{li2021degradation} requiring persistent long-term control over sensing hardware is currently out of scope but forms a key aspect of our future work.

\noindent \textbf{Privacy adversary}: The privacy adversary observes the publicly disclosed DP-perturbed covariance $\hat{S}_t$ and whitened residual $\hat{\tau}^{res}_t$ or corresponding functional artifacts of the hypothesis test such as test statistic estimates or levels of significance. The privacy-oriented adversarial goal is to attempt to infer the underlying raw sensor measurements $y_t$ or system state $x_t$. The privacy adversary is assumed to be computationally bounded and has knowledge of the DP mechanism and its parameters.

\noindent \textbf{Trust Assumptions}: 
We operate under a semi-honest trust model regarding the disclosure recipients which include regulatory agencies and public bodies. While these entities are assumed to faithfully execute the verification protocol using the disclosed statistics $\hat{S}_t$ (i.e., they perform valid statistical checks), we assume they are also curious about the underlying system dynamics. Consequently, they may attempt to exploit the disclosure to infer sensitive proprietary information such as raw sensor data $y_t$ or state space dynamics such as $x_t$ that exceeds their authorization. Therefore, the privacy guarantees must hold even against such curious recipients. Lastly, we assume that the disclosure channel between utility and regulators itself is trustworthy.

\noindent \textbf{Security Objective}: 
The primary security objective of PRICE is to ensure that public disclosures of the DP-perturbed covariance $\hat{S}_t$ and residual $\hat{\tau}^{res}_t$ do not compromise the confidentiality of the underlying raw sensor measurements $y_t$. We aim to satisfy formal Differential Privacy guarantees such that no curious observer, whether a third-party entity or the regulatory body, can distinguish the true underlying sensor data beyond the pre-agreed privacy budgets $(\epsilon_{cov}, \gamma_{cov})$ and $(\epsilon_r, \gamma_r)$. This ensures that stakeholder ICS dynamics remain hidden from unauthorized inference attempts while maintaining the integrity of the disclosed verification statistics.

\subsection{Differential Privacy Primer}
We provide a brief summary of the following useful concepts pertaining to differential privacy.
\begin{definition}\label{thmdp}\cite{dwork2014algorithmic} : A randomized mechanism $\mathcal{M}:\mathbb{R}^n\rightarrow\mathbb{R}^k$ is said to preserve $(\epsilon,\delta)$- differential privacy with respect for all points $x_1,x_2 \in \mathbb{R}^n$, the following holds 
\[
\mathbb{P}(\mathcal{M}(x_1)\in \mathcal{O})\leq e^{\epsilon}\mathbb{P}(\mathcal{M}(x_2)\in\mathcal{O}) + \delta
\]
\end{definition}
\noindent
In other words, Definition \ref{thmdp} ensures that for any two points $x_1,x_2$ in domain $\mathbb{R}^n$, the probability that mechanism $\mathcal{M}(x_1)$ leads to an output in the set $\mathcal{O} \subseteq \mathbb{R}^k$ is upper bounded by the probability that mechanism $\mathcal{M}(x_2)$ leads to an output in the set $\mathcal{O} \subseteq \mathbb{R}^k$ scaled by exponentiation of the privacy loss $\epsilon>0$ with an addition of $\delta>0$.

For DP mechanisms, we characterize the concept of adjacency with respect to a multi-variate function $f:\mathbb{R}^d\rightarrow \mathbb{R}^{d}$. Any two elements $x_1,x_2\in\mathbb{R}^d$ are deemed to be adjacent, if they differ in at most 1 element. Consequently, the sensitivity of a function depends on the maximum difference in function values caused by adjacent elements. Definition \ref{defn2} formalizes the concept of sensitivity.
\begin{definition}\label{defn2}:
The $l_k$-sensitivity of a $d$ dimensional function $f:\mathbb{R}^d\rightarrow \mathbb{R}^{d}$ is defined as $\Delta_{kf} = \underset{x,y}{max}||f(x)-f(y)||_k$, where $x$ and $y$ are adjacent elements.
\end{definition}
Most common DP formulations utilize $k=\{1,2\}$ denoting $l_1$ or $l_2$ sensitivity assumptions respectively. In this paper, we adopt a two phase approach towards differential privacy wherein each phase is applied sequentially and pertains to protecting the privacy of the covariance matrix and residual vectors respectively. More specifically, we leverage a DP-based method \cite{amin2019differentially} for privacy preserving covariance matrix factorization while employing a Gaussian Differential Privacy (GDP) approach for residual vectors. 
\begin{definition}\label{defn3}\cite{dwork2014algorithmic}
A mechanism is deemed to follow Gaussian Differential Privacy (GDP) if it injects independent noise $e_i \sim \mathcal{N}(0, \sigma^2)$, where $\sigma > 0$ is the noise scale, to each component of $p$ dimensional function $f: \mathbb{R}^p \rightarrow \mathbb{R}^p$. 
\end{definition}
The GDP mechanism provides $(\epsilon,\gamma_{r})$-DP when $\epsilon \in(0,1)$, $c^2>2ln(1.25/\gamma_{r})$ and $\sigma\geq c\Delta_2/\epsilon$. As a result, we have Equations \eqref{eq:gdp1} and \eqref{eq:gdp2}. Since the $l_2$ sensitivity $\Delta_2$ is used in the context of GDP of residuals, we use $\Delta_2=\Delta_r$ for notational clarity.
\begin{gather}
    \sigma > \frac{\Delta_r}{\epsilon}\sqrt{2ln(\frac{1.25}{\gamma_r})} \label{eq:gdp1}\\
    \mathbb{P}\Big(\Big|\sum\limits_{i=1}^{p} e_{i,t} \Big| \geq \theta_r \Big) \leq \gamma_r\text{ , } \theta_r = \frac{\sigma^2\epsilon}{\Delta_r} - \frac{p\Delta_r}{2} \label{eq:gdp2}
\end{gather}



\subsection{Privacy Preserving Covariance Disclosures}\label{subsec:ppcd}
We utilize the Laplacian noise based DP method \cite{amin2019differentially} that factorizes $S_t$ to yield a DP driven perturbed matrix $\hat{S_t}$ as given in Equation \eqref{eq:pce0a}. However, instead of directly perturbing $S_t$, we compute its square root factorization $\tilde{S}_t = V_t\Lambda^{\frac{-1}{2}}_tV_t^T$. Next we perturb eigenvalues of $\tilde{S}_t$ as specified in \cite{amin2019differentially} so as to compute $\hat{S}_t =\hat{V_t}\hat{\Lambda}\hat{V_t}^T $. Consequently, we obtain $\hat{S}^{\frac{-1}{2}}_t = \hat{V_t}\hat{\Lambda}^{\frac{-1}{2}}$. Using $\hat{S}^{\frac{-1}{2}}_t$, we can also obtain a perturbed lower dimensional residual $\hat{\tau}^{cov}_t$ and its corresponding test statistic $ \hat{T}^{cov}_{\chi^2,t}$ denoted by Equations \eqref{eq:pce0b} and \eqref{eq:pce0c} respectively. Using Laplacian noise to perturb eigenvalues as described in \cite{amin2019differentially} ensures $(\epsilon_{cov},\gamma_{cov})$-DP for all $\lambda_i, i \in \{1,d\}$ such that $\hat{\lambda}_i = \lambda_i + Lap(\Delta_{l}/\epsilon_{cov})$.
\begin{gather}
    \hat{S_t} = \hat{V_t}\hat{\Lambda}_t\hat{V_t}^T \label{eq:pce0a}\\
    \hat{\tau}^{cov}_t = (\hat{S_t})^{-1/2}r_t \label{eq:pce0b}\\
    \hat{T}^{cov}_{\chi^2,t} = (\hat{\tau}^{cov})^T_t\hat{\tau}^{cov}_t = ||\hat{\tau}^{cov}_t||^2_2 \label{eq:pce0c}
\end{gather}
Further, we note that the perturbed test statistic $\hat{T}^{cov}_{\chi^2,t}$ is computed using a DP-induced covariance matrix $\hat{S}_t$ \emph{and an unperturbed} residual vector $r_t$.
Consequently, we can state the relations defined by Equations \eqref{eq:pce2} and \eqref{eq:pce3} as stated in \cite{amin2019differentially}.
\begin{gather}
E_l = ||\Lambda-\hat{\Lambda}||_1 = \underset{i}{max} |\hat{\lambda}_i - \lambda_i|\label{eq:pce2}\\
\mathbb{P}\Big(E_l \leq \frac{\Delta_l}{\epsilon_{cov}}log\Big(\frac{d}{\gamma_{cov}}\Big) \Big) \geq 1-\gamma_{cov} \label{eq:pce3}
\end{gather}
Equation \eqref{eq:pce3} ensures that the maximum absolute Laplacian noise value given by $E_l$ as defined in Equation \eqref{eq:pce2} is less than $\frac{\Delta_l}{\epsilon_{cov}}log\Big(\frac{d}{\gamma_{cov}}\Big)$ with probability at least $1-\gamma_{cov}$. Analyzing the effect of DP-induced covariance matrices helps us establish Lemma \ref{thm1}.

\begin{lemma}\label{thm1}
The original and the perturbed test statistics $\hat{T}^{cov}_{\chi^2,t}$, $T_{\chi^2,t}$ satisfy the following relation for $p\leq d$:
\[ \mathbb{P}\Big[ | \hat{T}^{cov}_{\chi^2,t} - {T}_{\chi^2,t} | \leq  R_t\theta_l \Big] \geq 1-\gamma_{cov} \]
where $R_t = \sum\limits^{p}_{i=1} (r^i_t)^2$ and $\theta_l = \Big(\frac{\Delta_l}{\epsilon_{cov}}log\Big(\frac{d}{\gamma_{cov}}\Big)\Big)$
\end{lemma}
\begin{proof}
\noindent Proof provided in Appendix \ref{subsec:pf_thm1}.    
\end{proof}
Lemma \ref{thm1} enables us to probabilistically bound the observed difference in test statistic value, \emph{only when DP is applied on the covariance matrix}. However, in the interest of full compliance, it is also important to reveal the reduced residual values, such that the regulator can carry out the entire hypothesis test workflow represented in Equations \eqref{eq:nullh} - \eqref{eq:alarm_condition}. Doing so, would require privacy protections on residuals as well, which we explore in the following section.

\subsection{Privacy Preserving Residual Disclosure}\label{subsec:pprd}
We consider the unperturbed low dimensional residual representation $\tau_t = (S_t)^{-1/2}(r_t)$ where $\tau_t \in \mathbb{R}^p$ as defined in Section \ref{subsec:htad}. We use the GDP mechanism presented in Definition \ref{defn3} to perturb $\tau_t$ according to Equation \eqref{eq:dp_res}.
\begin{gather}
\hat{\tau}^{res}_t = \tau_t + e_t, e_t \sim N(0,\sigma^2I) \label{eq:dp_res}\\
\hat{T}^{res}_{\chi^2,t} = (\hat{\tau}^{res}_t)^T\hat{\tau}^{res}_t = ||\hat{\tau}^{res}_t||^2_2\label{eq:dp_res2}
\end{gather}
$\hat{T}^{res}_{\chi^2,t}$ in Equation \eqref{eq:dp_res2} represents the test statistic obtained \emph{purely} through the GDP perturbation of $\tau_t$. Generating $\hat{T}^{res}_{\chi^2,t}$ is especially useful in enabling implementation flexibilities for compliance verification methods as discussed in Section \ref{sec:ppaf}. The GDP perturbation leads us to Lemma \ref{lem1} which provides probabilistic bounds on the GDP noise vector $e_t$. 
\begin{lemma}\label{lem1}
Under conditions of GDP, for a given value of $\gamma_r,\epsilon$, the following condition must hold
\[ \mathbb{P}\Big[  ||e_t||^2_2 \leq \frac{\theta_r^2}{p}    \Big] \geq (1-\gamma_r)^p \]
\end{lemma}
\begin{proof}
\noindent Proof provided in Appendix \ref{subsec:pf_lem1}.    
\end{proof}
Lemma \ref{lem1} is vital towards deriving the overall privacy implications when the covariance as well as residual perturbations are integrated and presented to the regulator for compliance verification. 

We now turn our attention towards characterizing the probability distribution of the GDP induced test statistic $\hat{T}^{res}_{\chi^2,t}$. As a result, we consider the non-central $\chi^2$ distribution with non-centrality parameter $\mu$ and $k$ degrees of freedom denoted by $\chi^{2}(k,\mu)$. Further, we denote $F^{\tau_t}_{\chi^2,p}(x)$ as the CDF of $\chi^{2}(p,||\tau_t||^2_2)$. Consequently, we derive Lemma \ref{lem2} which establishes the probability distribution of the variance scaled test statistic $(\hat{T}^{res}_{\chi^2,t}/\sigma^2)$ under GDP provisions.
\begin{lemma}\label{lem2}
Under GDP, $(\frac{\hat{T}^{res}_{\chi^2,t}}{\sigma^2}) \sim \chi^{2}(p,\frac{||\tau_t||^2_2}{\sigma^2})$
\end{lemma}
\begin{proof}
\noindent Proof provided in Appendix \ref{subsec:pf_lem2}.    
\end{proof}
Lemma \ref{lem2} provides a distributional characterization of the perturbed residual test statistic under GDP conditions. Therefore, it serves as a precursor to Theorems \ref{thm2} helping establish probabilistic bounds between perturbed and original low dimensional residual representations. Lemma \ref{lem2} also proves to be an important enabler for computing DP-informed level of significance for carrying out the $\chi^2$ hypothesis test at the regulator as represented in Theorem \ref{thm3}.

\begin{theorem}\label{thm2}
For an $(\epsilon,\delta)$-DP Gaussian mechanism on a given $\tau_t$, the following result holds
\begin{align*}
{\mathbb{P}\Big[ L \leq (\hat{T}^{res}_{\chi^2,t} - T_{\chi^2,t}) \leq U \Big|||e_t||^2_2 \leq \frac{\theta_r^2}{p}\Big]}\\ \geq (F^{\tau_t}_{\chi^2,p}(U) - F^{\tau_t}_{\chi^2,p}(L)) (1-\gamma_{r})^p
\end{align*}
with $L = \frac{\theta_r}{\sigma^2p}\Big(\theta_r-2\sum\limits_{i=1}^{p}\tau_{i,t}\Big)$ and $U= \frac{\theta_r}{\sigma^2p}\Big(\theta_r+2\sum\limits_{i=1}^{p}\tau_{i,t}\Big)$
\end{theorem}
\begin{proof}
\noindent Proof provided in Appendix \ref{subsec:pf_thm2}.    
\end{proof}
The main contribution of Theorem \ref{thm2} lies in its ability to link the 
perturbed test statistic $\hat{T}^{res}_{\chi^2,t}$ as a function of the non-central $\chi^2$ CDF with $p$ degrees of freedom centered on the original test statistic $T_{\chi^2,t}$. As a consequence of Theorem \ref{thm2}, we can derive lower and upper bounds on the difference between original and perturbed tests statistics purely as a function of GDP parameters and the unperturbed low dimensional residual representation $\tau_t$. 


\subsection{Integrating Covariance and Residual Privacy}
Lemma \ref{thm1} deals with aspects of privacy when only the covariance undergoes DP, Theorem \ref{thm2} applies to a scenario wherein GDP is applied only to lower dimensional residual representations. However, integrating both of these individual DP steps is important so as to enable the disclosures of both covariance and residuals \emph{separately} by the utility. Separate disclosures are vital to recreate the hypothesis test workflow at the regulator level to help satisfy compliance while preserving privacy of utility level operations. In this section, we address two issues which arise as a consequence of separate disclosures. First, introduction of DP noise across the covariance and residual steps might provide different test results at the regulator compared those of the original test at the utility level. Second, minimizing these misclassifications at the regulator level requires an alternate version of the test characterized by a level of significance customized for DP noise injected at the covariance and residual levels. Lastly, regulator level tests that consider separate disclosures covariance and residuals require strong privacy bounds that can help inform the implementation modalities. 

We consider $T_{\chi^2,t}$, $\hat{T}^{cov}_{\chi^2,t}$ and $\hat{T}^{res}_{\chi^2,t}$ corresponding to $\tau_t$, $\hat{\tau}^{cov}_t$ and $\hat{\tau}^{res}_t = \hat{\tau}^{cov}_t + e_t$ respectively. Further, we consider 
\begin{gather}
\Pi_t = \{r_t, \gamma_{cov}, \gamma_{res}, \epsilon_{cov},\epsilon_{l},\epsilon_{r}, \Delta_l,\Delta_r\} \label{eq:pit}
\end{gather}
In Equation \eqref{eq:pit}, $\Pi_t$ denotes the set of observed residuals and DP parameters at time $t$. We let $\hat{\rho}_t, \rho_t$ denote the alarms raised with and without differential privacy respectively. We first establish Lemma 3 as a means to determine the upper bound the reduced dimensional DP residual $\tau$ given the distributional knowledge of the perturbed test statistic. 

\begin{lemma}\label{lem3}
Given $\hat{T}^{res}_{\chi^2,t} (\tau) = ||\tau + e||^2_2$, $e \sim N(0,\sigma^2I)$, 
\[\underset{\tau}{argmax} \text{ } \mathbb{P}\Big[ \hat{T}^{res}_{\chi^2,t} > \sigma^2\phi \Big] = \underset{\tau}{argmax}(||\tau||_2) \]  
where $\phi$ belongs to the support of distribution $\chi^{2}(p,\frac{||\tau||^2_2}{\sigma^2})$
\end{lemma}
\begin{proof}
\noindent Proof provided in Appendix \ref{subsec:pf_lem3}.    
\end{proof}

We now consider the integration of DP driven covariance matrix structure defined in Section \ref{subsec:ppcd} and the GDP induced residuals discussed in Section \ref{subsec:pprd}. In order to do so, we consider the sequential application of DP on the covariance matrix followed by the GDP phase on the residual. Adopting a sequential approach enables a seamless framework of theoretical analysis that can be used to derive privacy oriented misclassification rates, alternate levels of significance as well as efficient and customizable implementation strategies. As a result, we have Equations \eqref{eq:int1}-\eqref{eq:int2} that represent the \emph{sequential privacy scheme}.
\begin{gather}
\hat{\tau}^{cov}_t = (\hat{S_t})^{-1/2}r_t \label{eq:int1}\\
\hat{\tau}^{res}_t = \hat{\tau}^{cov}_t + e_t, e_t \sim N(0,\sigma^2I) \label{eq:int2}
\end{gather}
\noindent
Next we consider $\chi^{2,NC}_{\hat{\alpha}}$ which denotes the $\hat{\alpha}$ level upper quantile of the non-central $\chi^2$ distribution given by $\chi^{2}(p,\hat{T}^{cov}_{\chi^2,t})$, where $\hat{T}^{cov}_{\chi^2,t} = ||\hat{\tau}^{cov}_{t}||^2_2$ forms the covariance privacy induced test statistic defined in Equation \eqref{eq:pce0c}.
We also note that $\chi^{2,NC}_{\hat{\alpha}}$ denotes the $\hat{\alpha}$ level upper quantile representing the level of significance for the hypothesis test carried out with DP measures. Similarly, $\chi^{2}_{\alpha}$ represents the $\alpha$ level of significance for the non-DP hypothesis test. We obtain Theorem \ref{thm3} which considers a scenario wherein a regulator tries to recreate the hypothesis test on the basis of utility stakeholder disclosures that follow the sequential privacy scheme. 
\begin{theorem}\label{thm3}
Under the sequential privacy scheme, the Type-I error rate of the DP hypothesis test with $\hat{\alpha}$ level of significance is upper bounded by $\mathcal{E}^{max}_I(\hat{\alpha})$, where  
\begin{align*}
\mathcal{E}^{max}_I(\hat{\alpha}) 
&\leq \Big[1-F^{gamma}_{r_{max}}(R_t\theta_l)\Big]\Big[ 1- F^{\hat{\tau}^{cov}_t}_{\chi^2,p}
\Big(\sigma^2\chi^{2,NC}_{\hat{\alpha}}\Big)\Big]\\
&+ \Big[F^{ex}(\theta_l)\Big]^p\Big[ 1- F^{\hat{\tau}^{cov}_{max,t}}_{\chi^2,p}\Big(\sigma^2\chi^{2,NC}_{\hat{\alpha}}\Big)\Big]
\end{align*}
where $F^{ex}(.)$, $F^{gamma}_{r_{max}}(.)$ are the CDFs of $Exp(\frac{\epsilon_{cov}}{\Delta_l})$ and $Gamma(p,\frac{\epsilon_{cov}}{\Delta_lr^2_{max}})$ respectively.
\end{theorem}
\begin{proof}
\noindent Proof provided in Appendix \ref{subsec:pf_thm3}.    
\end{proof}
Theorem \ref{thm3} helps characterize the Type-I error rate $\mathcal{E}_I(\hat{\alpha})$ of the hypothesis test with $\hat{\alpha}$ level of significance carried out at the regulator. We can see that $\mathcal{E}_I(\hat{\alpha})$ is dependent on $\theta_l$  which consists of the covariance DP parameter $\gamma^{cov}$, covariance DP induced low dimensional residual $\hat{\tau}^{cov}_t$, its corresponding maxima $\hat{\tau}^{cov}_{max,t}$ computed according to Lemma \ref{lem3} as well as their CDFs $F^{\hat{\tau}^{cov}_t},F^{\hat{\tau}^{cov}_{max,t}}$ respectively. Additionally, we note that $\mathcal{E}^{max}_I({\hat{\alpha}})$ also incorporates the GDP variance parameter $\sigma$ as well. Therefore, Theorem \ref{thm3} lays the foundation for computing the Type 1-error rate of the sequential privacy scheme that is collectively influenced by the privacy measures at the covariance and residual steps. We specifically derive a Gamma function dependent upper bound in Theorem \ref{thm3} since it is known to provide tighter bounds when attempting to characterize tail-probabilities \cite{dwork2010boosting,balle2020hypothesis}. 

A fundamental implication of Theorem \ref{thm3} is that it helps guide the choice for $\hat{\alpha}$ depending on the privacy parameters chosen by the stakeholders which also influence their privacy loss. Usually, stakeholders have an established level of significance $\alpha$ depending on local detection benchmarks and domain expertise. Therefore, estimating the function $(\mathcal{E}^{max}_I)^{-1}(\alpha) = \hat{\alpha}$ provides an equivalent DP level of significance as a function of a pre-existing $\alpha$. The inverse function estimation ensures that the Type-I error of the DP hypothesis test is upper bounded by the non-DP test.

As a consequence, stakeholders can choose a perturbed level of significance \( \hat{\alpha} \) such that \( \alpha = \mathcal{E}^{\max}_I(\hat{\alpha}) \) which can be shared with the regulator. This choice of $\hat{\alpha}$ ensures that the regulator can reconstruct the hypothesis testing workflow using privacy-preserving disclosures of covariance and residuals, while still maintaining a worst-case Type~I error rate that does not exceed that of the unperturbed \( \chi^2 \) test at the stakeholder level. In our framework, given a fixed stakeholder level $\alpha$, we use the Monte-Carlo simulation method to estimate $\hat{\alpha}$.

We extend Theorem \ref{thm3} to determine bounds on misclassification rates of the hypothesis tests conducted with and without DP which is presented in Theorem \ref{thm4}. 

\begin{theorem}\label{thm4}
Given $\hat{\tau}_t^{cov},\hat{\tau}^{res}_t, \hat{\alpha}$, the misclassification rates can be given as 
\[ \mathbb{P}[\hat{\rho}_t=0|\rho_t=1] \leq \omega_1\Big[ 1-F^{\hat{\tau}^{cov}_t}_{\chi^2,p}\Big(\hat{T}_t\Big)\Big] +\omega_2 \Big[ 1- F^{\hat{\tau}^{cov}_{max,t}}_{\chi^2,p}
\Big(\hat{T}_t\Big)\Big]\]
\[ \mathbb{P}[\hat{\rho}_t=1|\rho_t=0] \leq F^{\hat{\tau}^{cov}_t}_{\chi^2,p}\Big(\hat{T}_t\Big)\Big[\omega_1+\omega_2\Big]\]
where,\\
$\omega_1 =\Big[1-F^{gamma}_{r_{max},\epsilon_{cov},\Delta_l}(R_t\theta_l)\Big]$, $\omega_2 = \Big[F^{ex}_{\Delta_l,\epsilon_{cov}}(\theta_l)\Big]^p$ and
$\hat{T}_t = T_{\chi^2,t} + \sigma^2\chi^{2,NC}_{\hat{\alpha}} - \chi^{2}_{\alpha}$
\end{theorem}
\begin{proof}
\noindent Proof provided in Appendix \ref{subsec:pf_thm4}.    
\end{proof}
Theorem \ref{thm4} formally states the misclassification rates that can occur with respect to the stakeholder and regulator purely on account of privacy preserving disclosures of covariance and residuals as part of the sequential privacy scheme. Using Theorem \ref{thm4}, we can see that when Gamma CDF values decrease and are more sensitive when $R_t$ is small, resulting in tighter bounds on misclassification rates. On the other hand, with larger $R_t$ values, Gamma CDF increases culminating in lower misclassification rates.

In addition to Theorem \ref{thm4}, we introduce a GDP noise calibration factor $\mu_t=||\hat{\tau}^{cov}_{max,t}||^2_2/\chi^{2,NC}_{\hat{\alpha}}$ to adjust the GDP noise variance $\sigma^2_t=\mu_t.\sigma^{min}$. The calibration factor is designed to incentivize more targeted application of noise for computing residual disclosures. This is particularly useful in high residual cases (such as during an attack window) where the DP threshold $\chi^{2,NC}_{\hat{\alpha}}$ might fail to deliver a good detection rate. As a result, we leverage $\mu_t$ to improve the power of the test in a dynamic, DP friendly fashion.


\section{DP driven Algorithmic Framework}\label{sec:ppaf}
We delineate our proposed algorithmic framework into two distinct components pertaining to the regulatory bodies and the utility stakeholders. For the utility level component, we focus on the development of a detection framework as well as relevant data disclosures based on residuals observed from the non-linear Kalman Filter model. On the other hand, the regulatory component purely focuses on the verification aspects based on disclosed data. Specifically, we consider a set of $j\in \mathcal{J}$ utilities, where $|\mathcal{J}| = J$. We divide the time horizon into discrete time steps denoted by $t$ that yield a distinct observation of sensor measurements as well as its associated residual at each utility. Further, we group these time steps into sets of evaluation epochs $w$, with each epoch consisting of $W$ consecutive, discrete time steps. 

As a consequence of the guarantees derived in Section \ref{sec:ssm}, we can derive two distinct implementation modes of our algorithmic framework pertaining to critical region based compliance and P-value driven verification. \emph{It is important to note that both these implementation modalities are mutually exclusive and must be pre-determined with consensus among utilities and regulators}. We present the algorithmic framework for each implementation mode.

\subsection{Critical Region Based Verification}
For the critical region (CR) compliance verification, we assume that the sole regulatory objective is to verify alarms with respect to differentially private disclosures of residuals and covariance from utilities. Alarm verification can be done by the disclosure of the test statistic by the utility followed by the critical region threshold. 

\subsubsection{CR based Utility Level Detection} 
The CR based utility level detection framework can be described on the basis of Algorithm \ref{alg:dpalg1_ul}. In Algorithm \ref{alg:dpalg1_ul}, at each time step the utility observes residual values $r_t$ and $C_t$. This is followed by the computation of the aggregated residual $r_w$ and covariance matrix $S_w$ and the epoch alarm $\rho_w$ for the evaluation epoch $w$. Based on these quantities, the utility can compute DP driven disclosures of $\hat{\tau}^{res}_w$ and $(\hat{S}_w)^{-1/2}$. Finally, the utility transmits an information tuple $\Pi_w$ consisting of the perturbed covariance matrix $(\hat{S}_w)$ as well as the transformed DP perturbed residual $\hat{\tau}^{rg}_w$, the critical region threshold $\chi^{2,NC}_{\hat{\alpha}}$ and the detected alarm $\rho_w$ with the regulator. In addition, using the concept of post-processing immunity and composition \cite{dwork2014algorithmic}, we can state that the disclosure of $\hat{\tau}^{rg}_w$ preserves $(\epsilon,\delta)$ privacy as well. 
\begin{algorithm}[htbp]
 \caption{Utility Level CR Verification Algorithm}\label{alg:dpalg1_ul}   
    \begin{algorithmic}
        \For{w=0,1,2,\ldots }
        \For{t=0,1,2,\ldots W}
        \State observe $r_t$ and $C_t$ using NLKF model $\mathcal{K}$
        \EndFor
        \State compute $r_w = \sum\limits_{t=0}^{W}r_t$ and $S_w = \sum\limits_{t=0}^{W}S_t$
        \State compute $\rho_w$ based on Equations \eqref{eq:alarm_condition} 
        \State compute $(\hat{S}_w)^{-1/2}$ using Equation \eqref{eq:pce0a}
        \State compute $\hat{\tau}^{cov}$ using Equation \eqref{eq:pce0b}
        \State compute $\hat{\tau}^{res}_w =  \hat{\tau}^{cov}_w + e_w$ using Equation \eqref{eq:dp_res}
        \State compute $\hat{\tau}^{rg}_w = r_w + (\hat{S}_w)^{1/2}e_w$ using $\hat{\tau}^{res}_w$
        \State compute $\hat{\alpha}, \chi^{2,NC}_{\hat{\alpha}}$ using Theorem \ref{thm3}.
        \State transmit $\Pi_w = \Big[\hat{S}_w,\hat{\tau}^{rg}_w,\chi^{2,NC}_{\hat{\alpha}},\rho_w \Big]$ with regulator.
        \EndFor
    \end{algorithmic}
\end{algorithm}

\subsubsection{CR based Regulatory Level Verification}
Algorithm \ref{alg:dpalg1_rl} captures the sequence of steps taken by a regulator for critical region based verification. At the regulatory level, we consider the set of utilities given by $\mathcal{J}$. The regulator receives DP-based information tuple $\Pi^j_w$ for each utility $j\in \mathcal{J}$ at each evaluation epoch. Using information contained in $\Pi^j_w$, the regulator can compute the factorization of the DP based covariance matrix $\hat{C}^{-1/2}$ as well as obtain an estimate of the DP-driven test statistic $\hat{T}^{j,res}_w$. The regulator can compute an alarm depending on the value of the test statistic $\hat{T}^{j,res}_w$ and the value of the CR threshold $\chi^{j,2,NC}_{\hat{\alpha}}$ based on the conditions given by \eqref{eq:dp_alarm_condition}. 
\begin{equation}
\hat{\rho}^j_w =
\begin{cases}
1, & \text{if } \hat{T}^{j,res}_w > \chi^{j,2,NC}_{\hat{\alpha}}, \\
0, & \text{otherwise.}
\end{cases}
\label{eq:dp_alarm_condition}
\end{equation}
\begin{algorithm}[htbp]
 \caption{Regulator Level CR Verification Algorithm}\label{alg:dpalg1_rl}   
    \begin{algorithmic}
        \For{w=0,1,2,\ldots }
        \For{j=0,1,2 \ldots J}
        \State receive $\Pi^j_w$ from utility stakeholder $j$
        \State factorize $\hat{S}^j_w = \hat{V}^j_w\hat{\lambda}^j_w(\hat{V}^j_w)^T$ compute $(\hat{S}^j_w)^{-1/2}$
        \State compute $\hat{\tau}^{j,res}_w = (\hat{S}^j_w)^{-1/2}\hat{\tau}^{j,rg}_w$
        \State compute $\hat{T}^{j,res}_w = ||\hat{\tau}^{j,res}_w||^2_2$
        \State compute $\hat{\rho}^j_w$ based on Equation \eqref{eq:dp_alarm_condition}
        \State verify if $\hat{\rho}^j_w = \rho^j_w$
        \EndFor
        \EndFor
    \end{algorithmic}
\end{algorithm}
As a consequence of Algorithm \ref{alg:dpalg1_rl}, the regulator independently obtains an estimate of $\hat{\rho}^j_w$ which can be compared with the reported alarm $\rho^j_w$. The misclassification rates pertaining to $\hat{\rho}^j_w$ and $\rho^j_w$ is provided using Theorem \ref{thm4}. Additionally, we provide a probabilistic bound on the worst case privacy loss incurred as a consequence of the CR based verification mode captured in Algorithms \ref{alg:dpalg1_rl} and \ref{alg:dpalg1_ul} in Theorem \ref{thm5}. 
\begin{theorem}\label{thm5}
The disclosure of $\hat{\tau}^{rg}_w$ incurs a worst case privacy loss $\epsilon'$ with the following probabilistic bounds
\[
\mathbb{P}\Big[ \epsilon' \geq \mathcal{L}(\Delta_r,\sigma^2,\hat{S}_w) \Big] \leq 1-(1-\gamma_r)^p
\]
where $\mathcal{L}(\Delta_r,\sigma^2,\hat{S}_w) = \frac{\Delta_r}{\sigma^2}(\mathbf{1}^T\hat{S}^{-1}_w\mathbf{1})^2\Big(\frac{\theta^2_r}{p}+\frac{1}{2.\mathbf{1}^T\hat{S}^{-1}_w\mathbf{1}}\Big)$
\end{theorem}
\begin{proof}
Proof provided in Appendix \ref{subsec:pf_thm5}   
\end{proof}
Theorem \ref{thm5} tells us that if GDP failure risk is low, as indicated by $\gamma_r \rightarrow 0$, the likelihood of the overall privacy loss exceeding $\mathcal{L}(\Delta_r,\sigma^2,\hat{S}_w)$ will be negligible. Additionally, lower values of $l_2$ sensitivity of residuals (denoted by $\Delta_r$) and high GDP noise (denoted by variance $\sigma^2$) minimize the lower bound on the worst case privacy loss. Lastly, $\mathbf{1}^T\hat{S}^{-1}_w\mathbf{1}$ can be viewed as a scaled Rayleigh quotient for $\hat{S}^{-1}_w$ computed using the vector $\mathbf{1}$. We can observe generally that increasing covariance privacy noise, characterized by increasing $\Delta_l/\epsilon_{cov}$, leads to a diminished value of the scaled Rayleigh quotient implying a lower worst case privacy loss. 


\subsection{P-value Based Compliance}
For the P-value (PV) compliance verification, the regulatory objective is to ensure that alarms have been computed with the correct P-value at the utility level. In this case, the utility discloses the test statistic distribution parameters as well as the DP-equivalent level of significance for verification of alarms at the regulator level.

\subsubsection{PV based Utility Level Detection}
To facilitate PV compliance, the utility level detection algorithm is represented by Algorithm \ref{alg:dpalg3_ul}. Similar to a CR driven setting, the utility computes the values of $\hat{\tau}^{cov}_w,\hat{\tau}^{res}_w, \hat{\alpha}$ and generates an alarm $\rho_w$. It discloses the information tuple $\Pi_w$ consisting of the values $\hat{T}^{res}_w, \hat{T}^{cov}_w, \hat{\alpha}_w,\rho_w$ to the regulator.
\begin{algorithm}[htbp]
 \caption{Utility Level PV Verification Algorithm}\label{alg:dpalg3_ul}   
    \begin{algorithmic}
        \For{w=0,1,2,\ldots }
        \For{t=0,1,2,\ldots W}
        \State observe $r_t$ and $C_t$ using NLKF model $\mathcal{K}$
        \EndFor
        \State compute $r_w = \sum\limits_{t=0}^{W}r_t$ and $S_w = \sum\limits_{t=0}^{W}C_t$
        \State compute $\rho_w$ based on Equations \eqref{eq:alarm_condition} 
        \State compute $(\hat{S}_w)^{-1/2}$ using Equation \eqref{eq:pce0a}
        \State compute $\hat{\tau}^{cov}_w$ using Equation \eqref{eq:pce0b}
        \State compute $\hat{\tau}^{res}_w =  \hat{\tau}^{cov}_w + e_w$ using Equation \eqref{eq:dp_res}
        \State compute $\hat{T}^{res}_w = ||\hat{\tau}^{res}_w/\sigma||^2_2$ and $\hat{T}^{cov}_w = ||\hat{\tau}^{cov}_w/\sigma||^2_2$
        \State transmit $\Pi_w = \Big[\hat{T}^{res}_w, \hat{T}^{cov}_w, \hat{\alpha}_w,\rho_w \Big]$ to regulator.
        \EndFor
    \end{algorithmic}
\end{algorithm}

\subsubsection{PV based Regulator Level Detection}
The regulator level algorithm for PV compliance verification is given in Algorithm \ref{alg:dpalg4_rl}. The objective of the regulator in this case is to estimate the non-central chi-square distribution using $\hat{T}^{j,cov}_w$ as the centrality parameter according to \ref{lem2} which can be used to estimate $\chi^{j,2,NC}_{\hat{\alpha}_w}$. 
On the basis of the alarm condition represented by Equation \eqref{eq:dp_alarm_condition}, the regulator can independently obtain and validate the alarm $\hat{\rho}^j_w$ with respect to $\rho^j_w$ for each utility. 
\begin{algorithm}[htbp]
 \caption{Regulator Level PV Verification Algorithm}\label{alg:dpalg4_rl}   
    \begin{algorithmic}
        \For{w=0,1,2,\ldots }
        \For{j=0,1,2 \ldots J}
        \State receive $\Pi^j_w$ from utility stakeholder $j$ 
        \State obtain $\hat{\alpha}_w$, $\hat{T}^{j,res}_w$ and $\hat{T}^{j,cov}_w$ from $\Pi^j_w$
        \State use $\hat{T}^{j,cov}_w$ to compute $\chi^{j,2,NC}_{\hat{\alpha}_w}$ using Theorem \ref{thm3}.
        \State compute $\hat{\rho}^j_w$ based on Equation \eqref{eq:dp_alarm_condition}
        \State verify if $\hat{\rho}^j_w = \rho^j_w$
        \EndFor
        \EndFor
    \end{algorithmic}
\end{algorithm}
In the PV implementation mode the regulator has access to the parametrized probability distribution of the DP test statistic $\hat{T}^{j,res}_w$ denoted by the non-central chi-squared distribution with centrality parameter $\hat{T}^{j,cov}_w$. This enables the regulator to obtain P-value of the DP test statistic based on the perturbed level of significance $\hat{\alpha}$ divulged by the utility. Additionally, we can derive bounds on the privacy loss incurred through the disclosure of $\hat{T}^{j,cov}_w$ in Theorem \ref{thm6}.

\begin{theorem}\label{thm6}
The disclosure of $\hat{T}^{cov}_w$ results in an $(\epsilon',\delta')$ DP mechanism where
\begin{equation}
\begin{aligned}
\epsilon'\geq & \epsilon_{cov} + \frac{\Delta_r^TC^{-1}\Delta_r}{2\sigma^2}\\
\delta'\leq & \Phi\Big(\frac{\sigma^2(\epsilon'-\epsilon_{cov})}{||\Delta_r^TC^{-1}||}-\frac{\Delta_r^TC^{-1}\Delta_r}{2||\Delta_r^TC^{-1}||} \Big)\\
-& \Phi\Big(-\frac{\sigma^2(\epsilon'-\epsilon_{cov})}{||\Delta_r^TC^{-1}||}+\frac{\Delta_r^TC^{-1}\Delta_r}{2||\Delta_r^TC^{-1}||}\Big)
\end{aligned}
\end{equation}
where $\Phi$ denotes the CDF of $N(0,\Delta_r^TC^{-1}\Delta_r)$ 
\end{theorem}
\begin{proof}
Proof provided in Appendix \ref{subsec:pf_thm6}   
\end{proof}

Theorem \ref{thm6} provides several insights into the privacy implications regarding the disclosure of $\hat{T}^{cov}_w$. First, we observe that with higher GDP covariance ($\sigma^2$) and lower sensitivity ($\Delta_r$) individually contribute to $\epsilon'$ making it closer to $\epsilon_{cov}$. Additionally, as lower bound on $\epsilon'$ approaches $\epsilon_{cov}$, we can also observe that the privacy failure probability $\delta'$ also tends towards 0. These observations imply that a higher GDP noise covariance and lower sensitivities while disclosing $\hat{T}^{cov}_w$ increasingly tends towards an $(\epsilon_{cov},\gamma_{cov})$-DP paradigm.

\section{Experimental Results}
We primarily leverage the HAI Security dataset \cite{hai}, as well as the ORNL power system (ORNL-PS) attack dataset \cite{pan2015classification,pan2015developing} for carrying out all experiments. We provide additional information pertaining to the use of these datasets in Appendix \ref{subsec:ddm}. Additionally, we provide an overview of the system implementation details in Appendix \ref{subsec:sys_det}. We provide supplementary results of our experimental framework driven by a container based implementation in Appendix \ref{subsec:sys_imp}

\begin{figure}[!htb]
    \centering
    \subfigure[$\epsilon_{cov}$]{\includegraphics[width=0.245\textwidth,keepaspectratio]{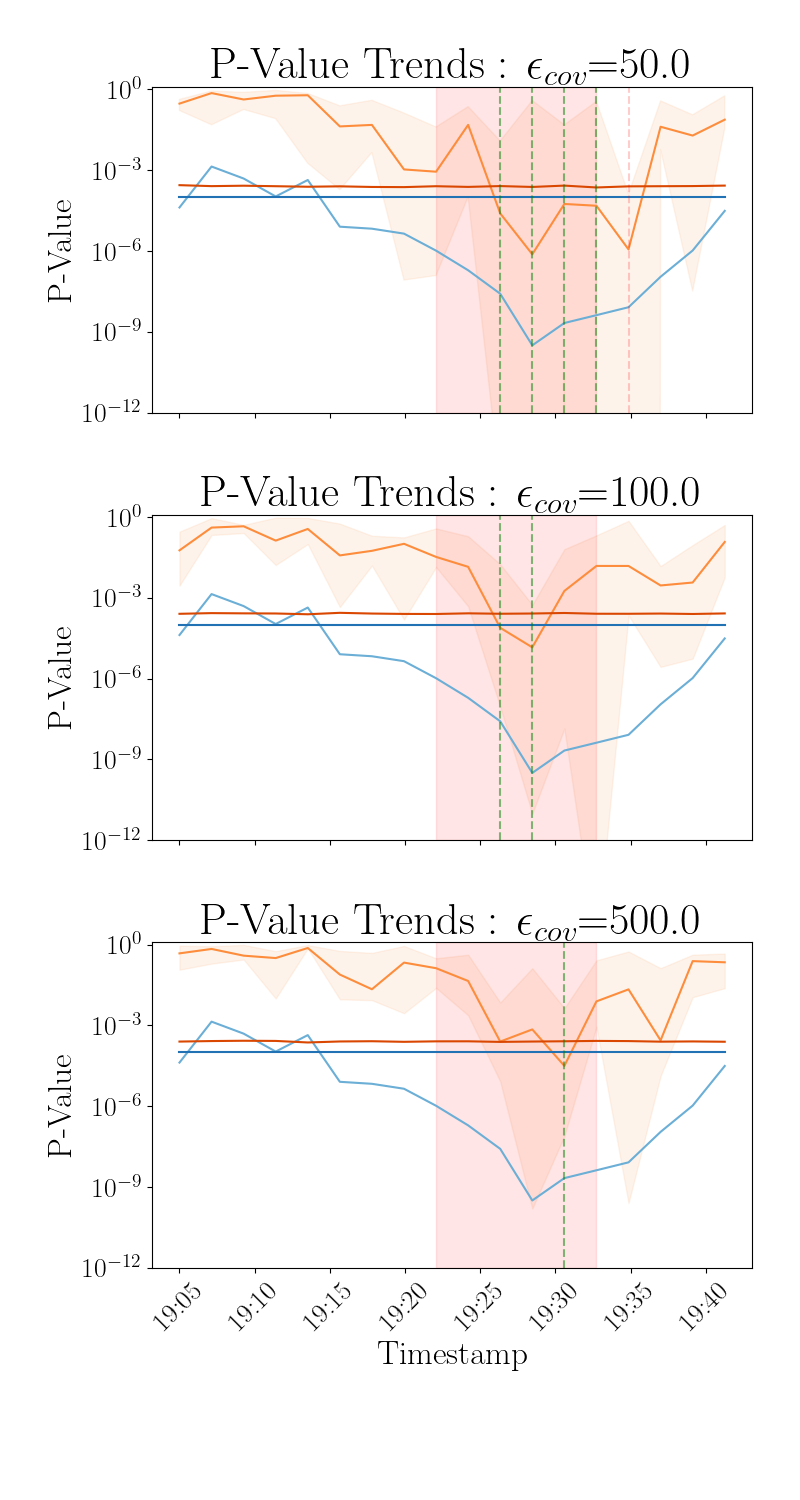}\label{fig:P1_pv_ec}}
    \hfill
    \subfigure[$\epsilon_{r}$]{\includegraphics[width=0.245\textwidth,keepaspectratio]{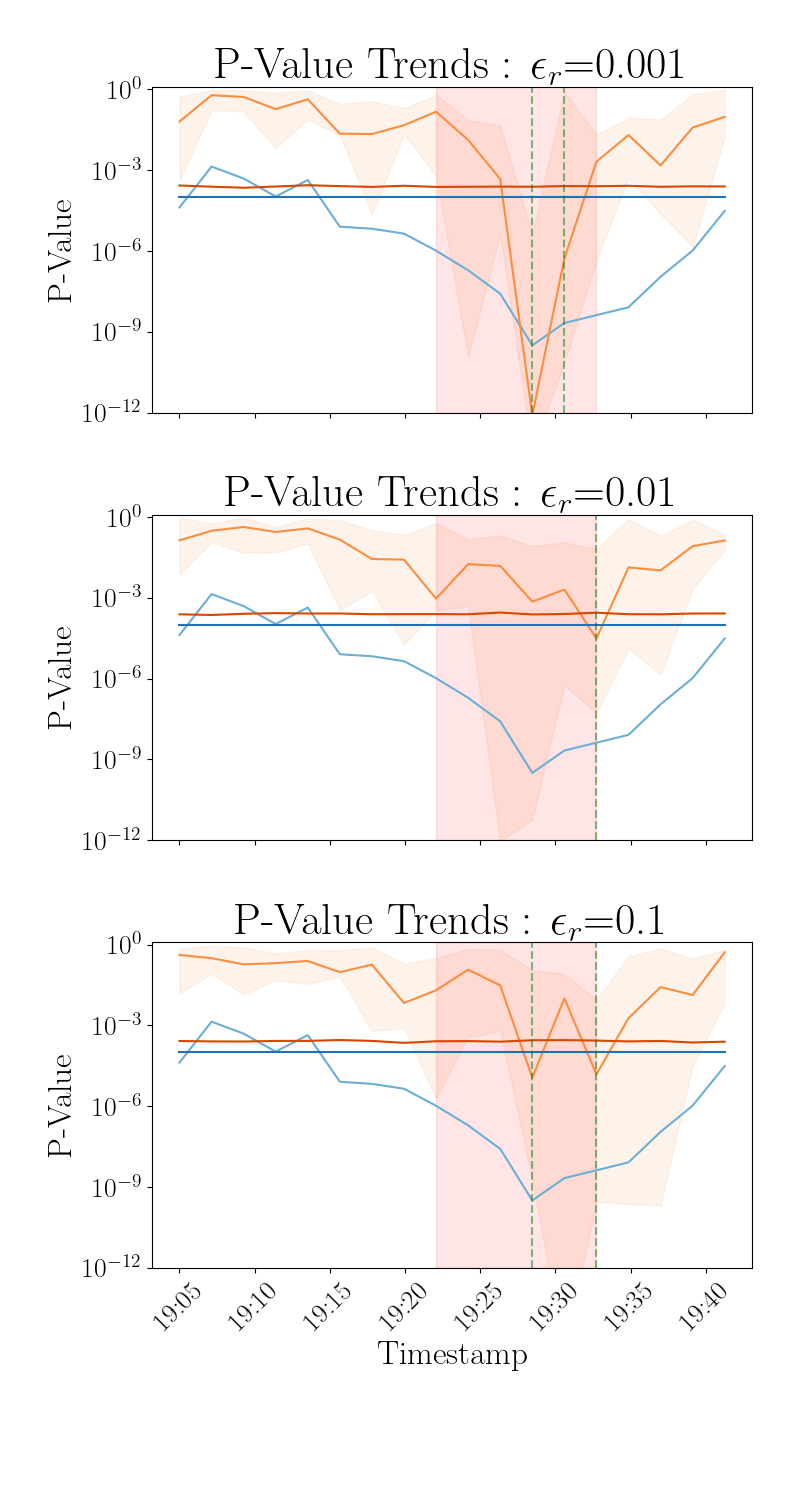}\label{fig:P1_pv_er}}
    \hfill
    \subfigure[$\gamma_{cov}$]{\includegraphics[width=0.245\textwidth,keepaspectratio]{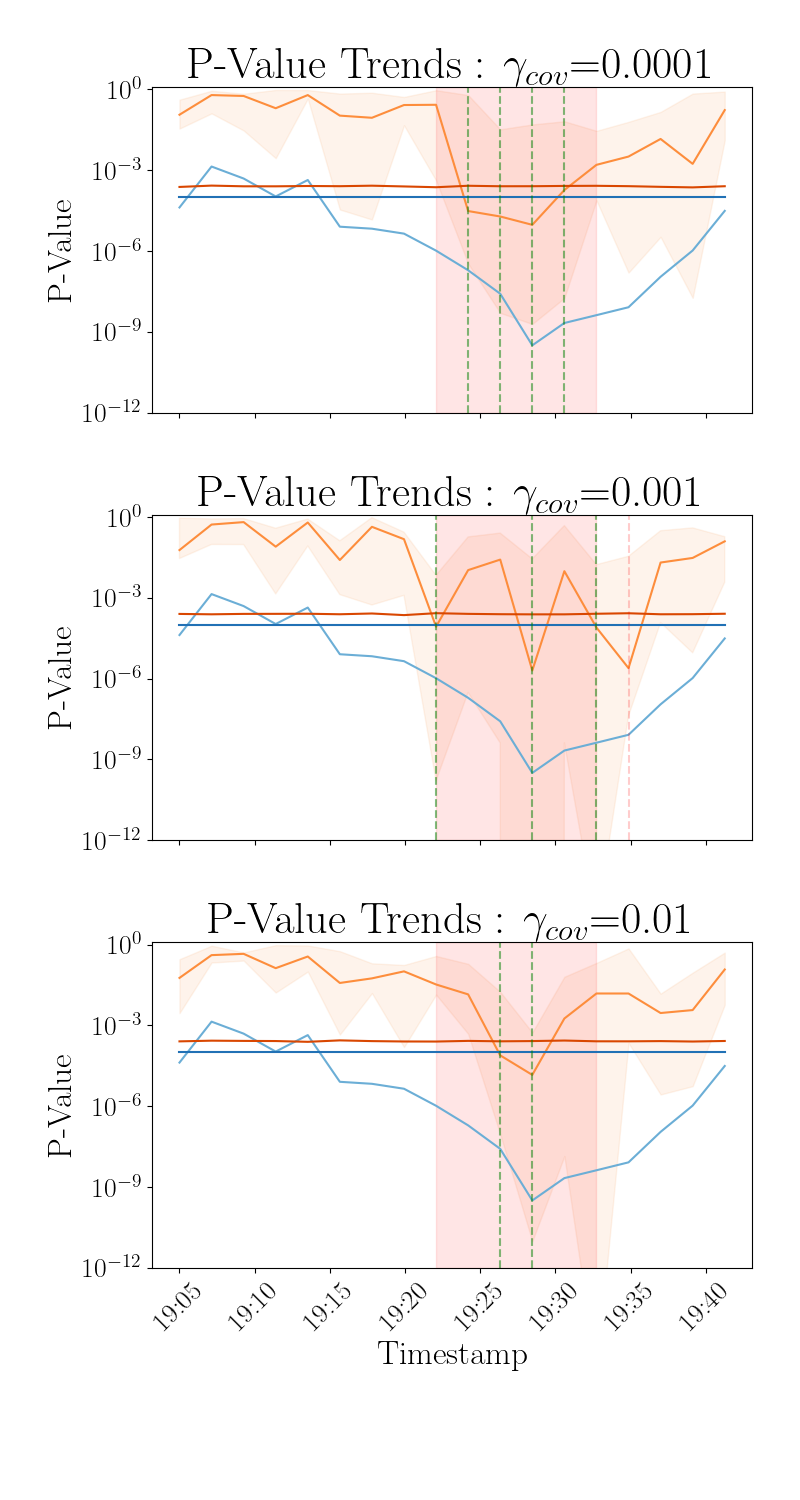}\label{fig:P1_pv_gc}}
    \hfill
    \subfigure[$\gamma_{r}$]{\includegraphics[width=0.245\textwidth,keepaspectratio]{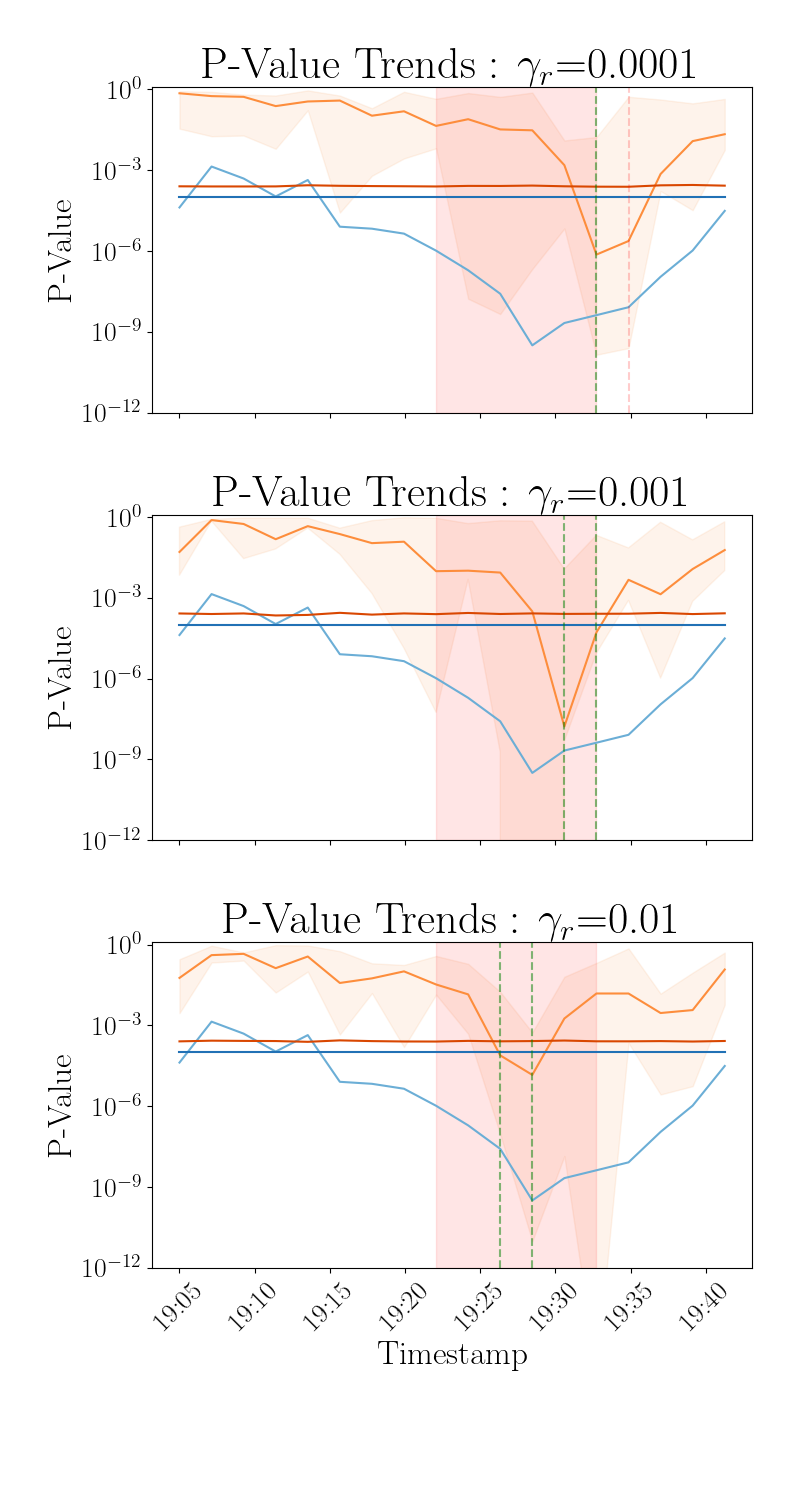}\label{fig:P1_pv_gr}}
    
    \vspace{-3.0em}
    \includegraphics[width=0.9\textwidth]{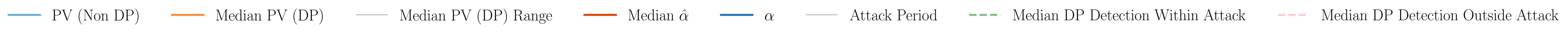}
    \vspace{3.0em}
    
    \caption{HAI Dataset: P-value trends for varying DP parameter values}
    \label{fig:P1_pv}
\vspace{-5mm}
\end{figure}
\begin{figure}[!htb]
    \centering
        \subfigure[$\epsilon_{cov}$]{\includegraphics[width=0.245\textwidth,keepaspectratio]{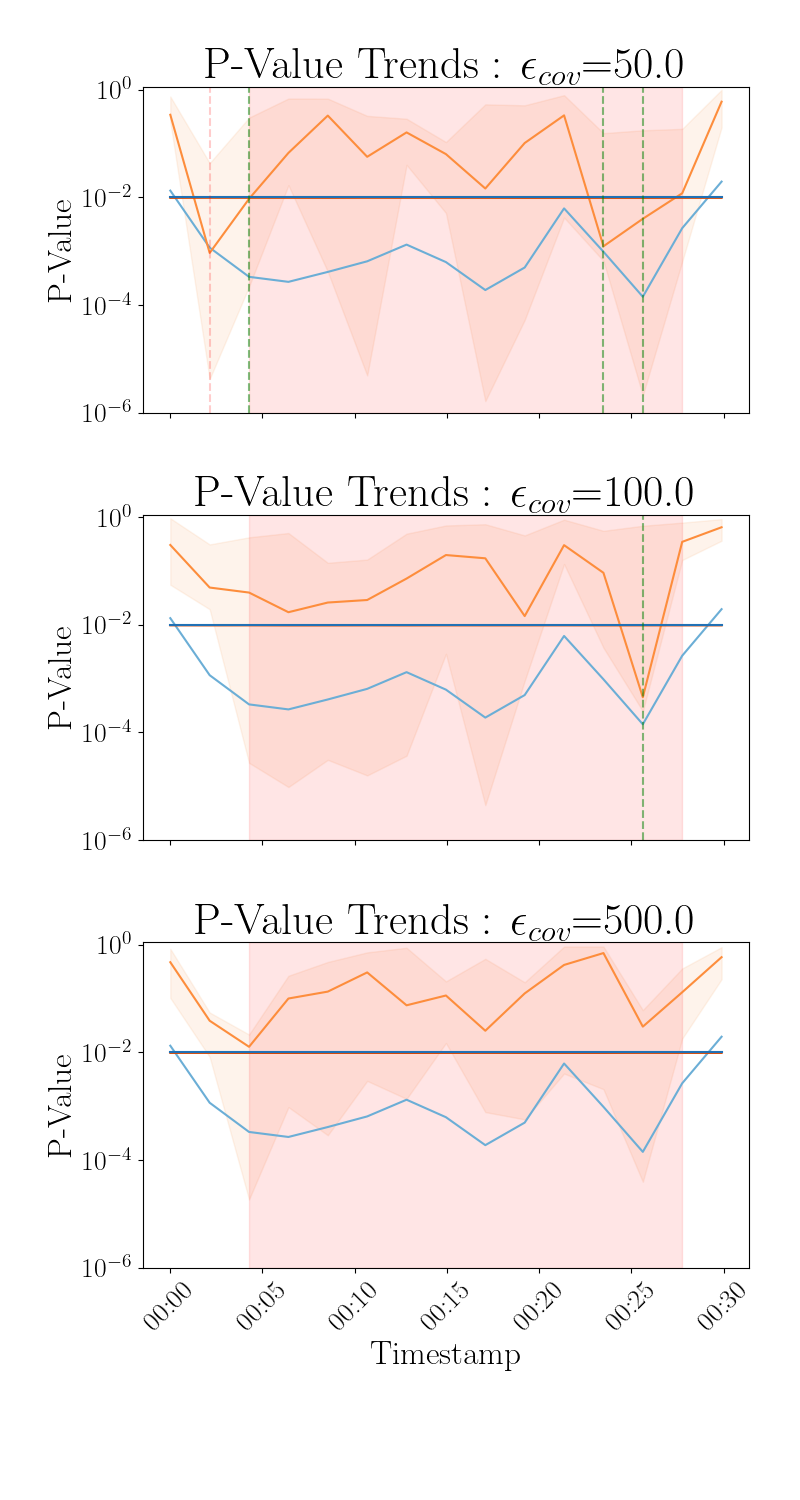}\label{fig:ps_pv_ec}}
        \hfill
        \subfigure[$\epsilon_{r}$]{\includegraphics[width=0.245\textwidth,keepaspectratio]{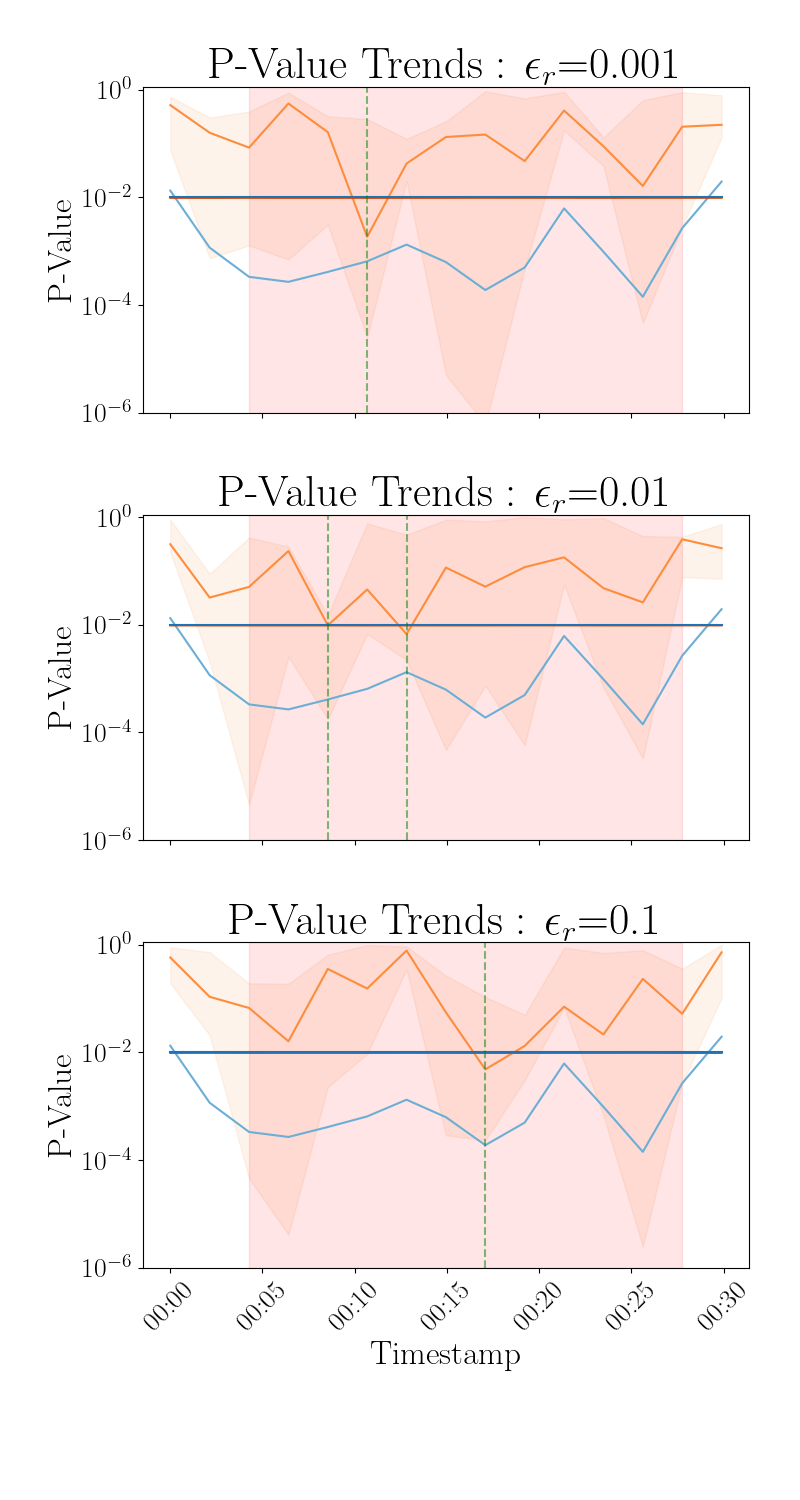}\label{fig:ps_pv_er}}
        \hfill
        \subfigure[$\gamma_{cov}$]{\includegraphics[width=0.245\textwidth,keepaspectratio]{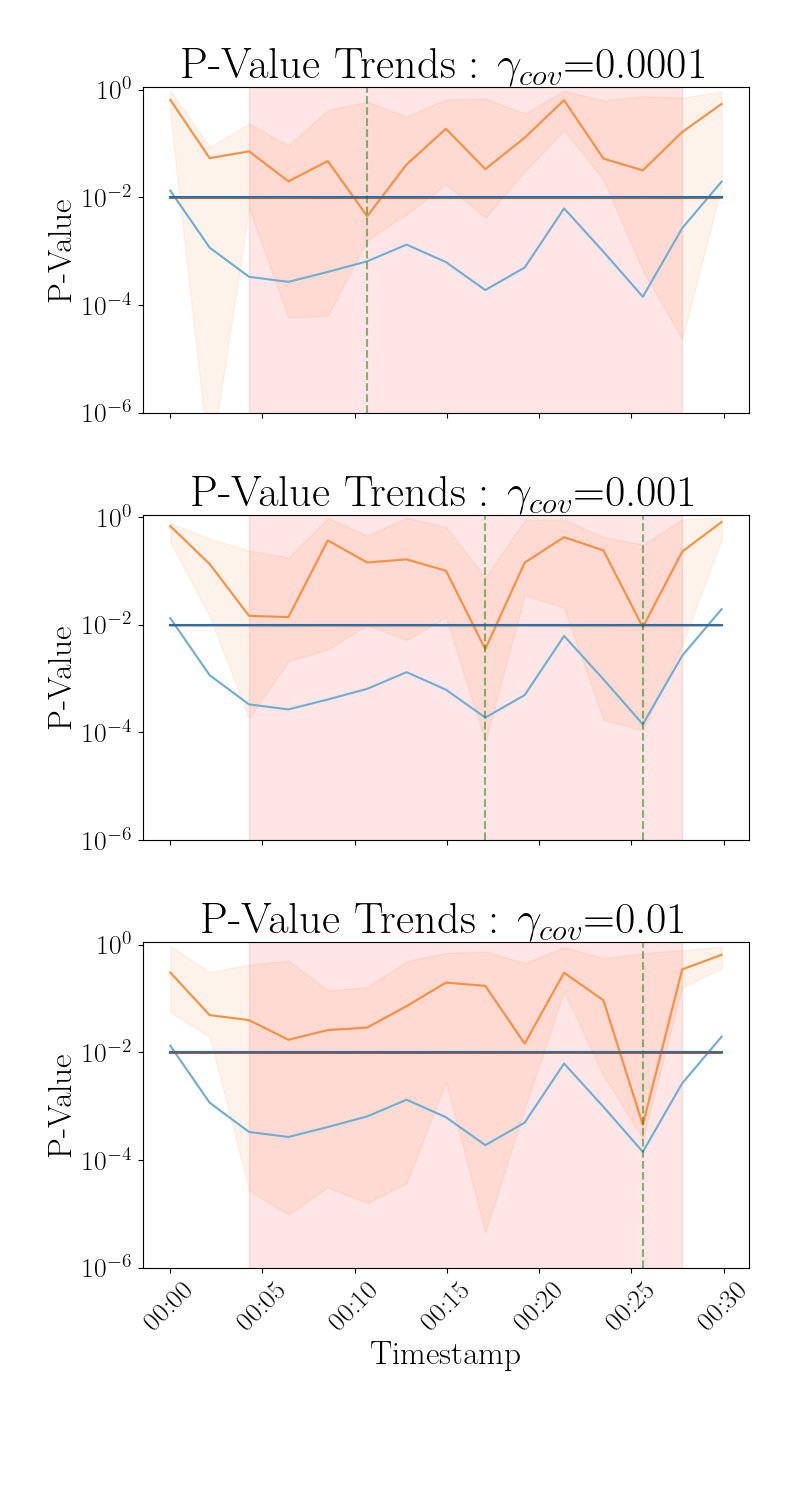}\label{fig:ps_pv_gc}}
        \hfill
        \subfigure[$\gamma_{r}$]{\includegraphics[width=0.245\textwidth,keepaspectratio]{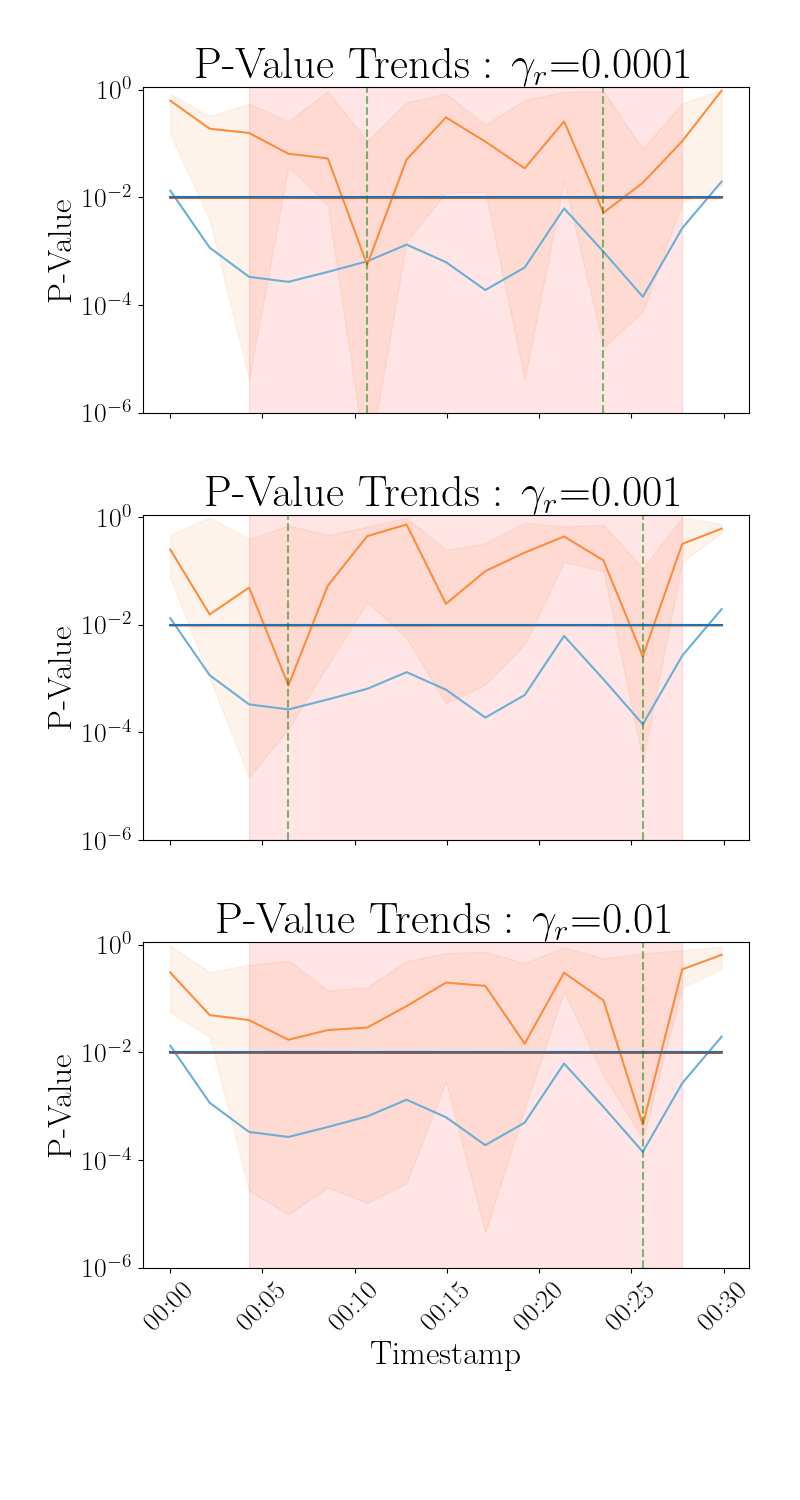}\label{fig:ps_pv_gr}}

        \vspace{-3.0em}
        \includegraphics[width=0.9\textwidth]{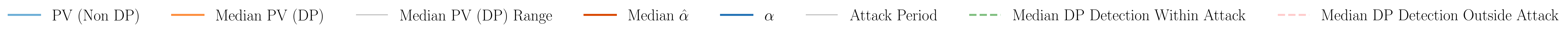}
        \vspace{3.0em}

  \caption{ORNL-PS Dataset: P-value trends for varying DP parameter values}\label{fig:ps_pv}
\vspace{-5mm}
\end{figure}

\begin{figure}[!htb]
    \centering
        \subfigure[$\epsilon_{cov}$]{\includegraphics[width=0.245\textwidth,keepaspectratio]{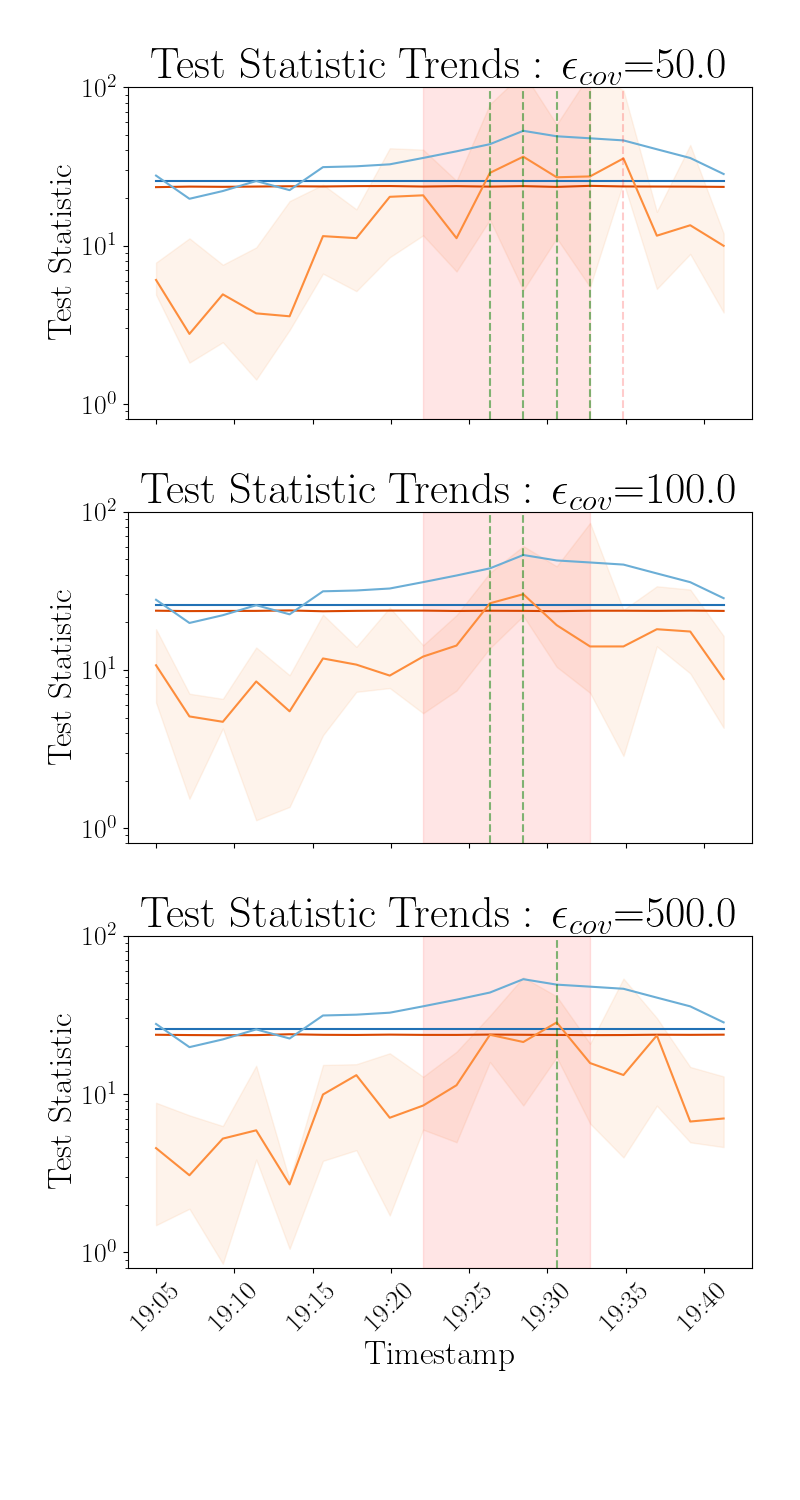}\label{fig:P1_ts_ec}}
        \hfill
        \subfigure[$\epsilon_{r}$]{\includegraphics[width=0.245\textwidth,keepaspectratio]{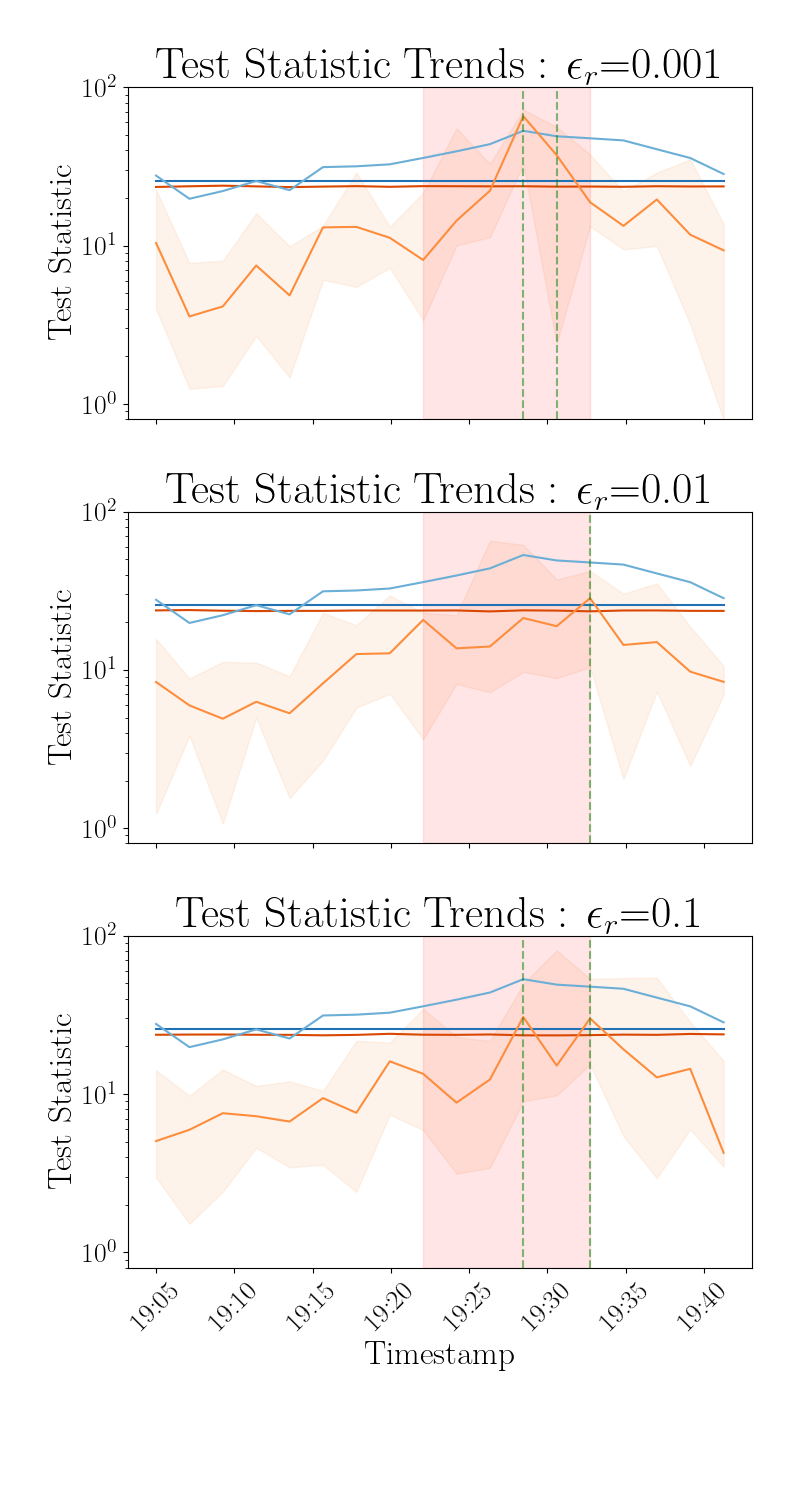}\label{fig:P1_ts_er}}
        \hfill
        \subfigure[$\gamma_{cov}$]{\includegraphics[width=0.245\textwidth,keepaspectratio]{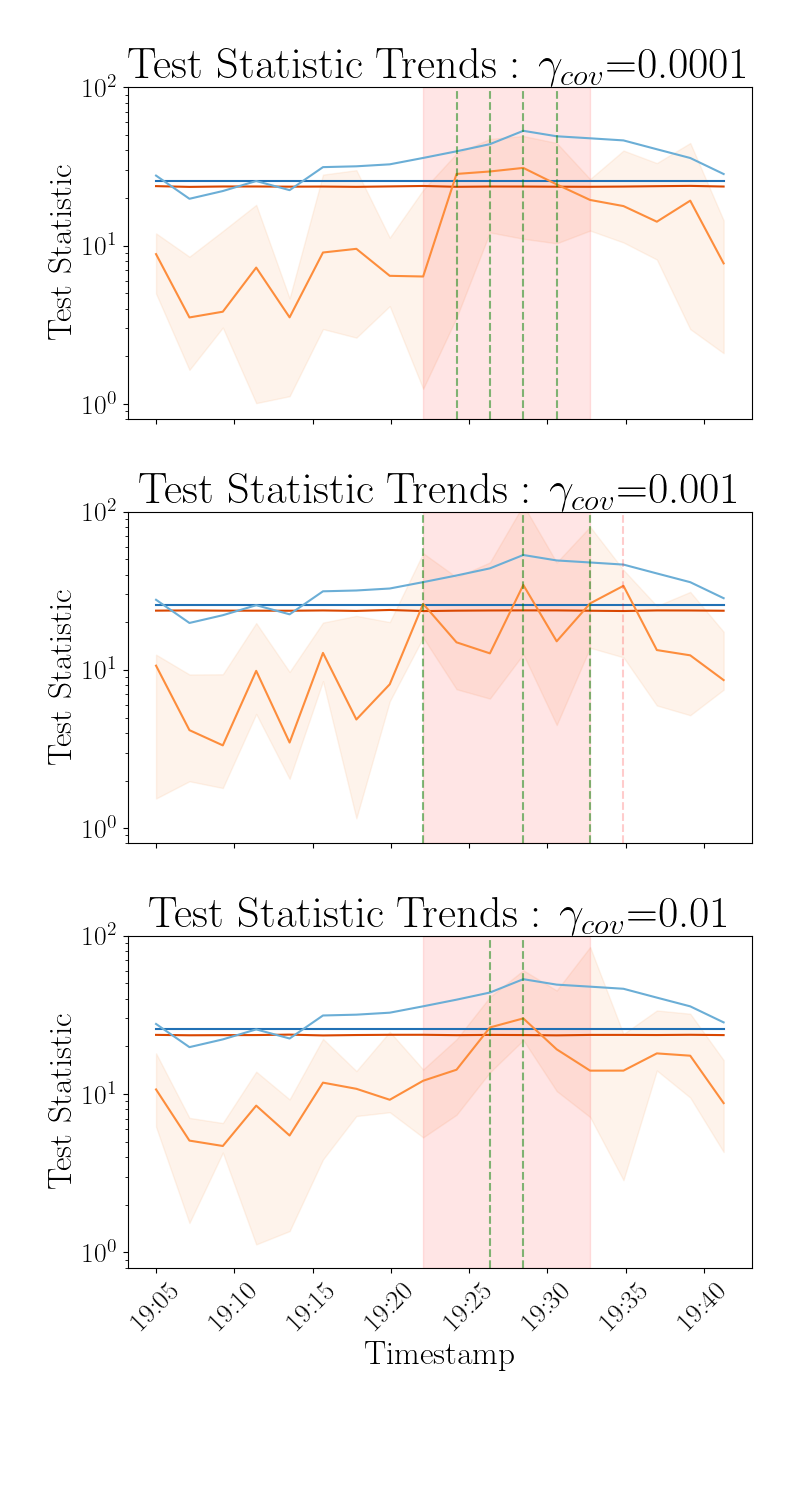}\label{fig:P1_ts_gc}}
        \hfill
        \subfigure[$\gamma_{r}$]{\includegraphics[width=0.245\textwidth,keepaspectratio]{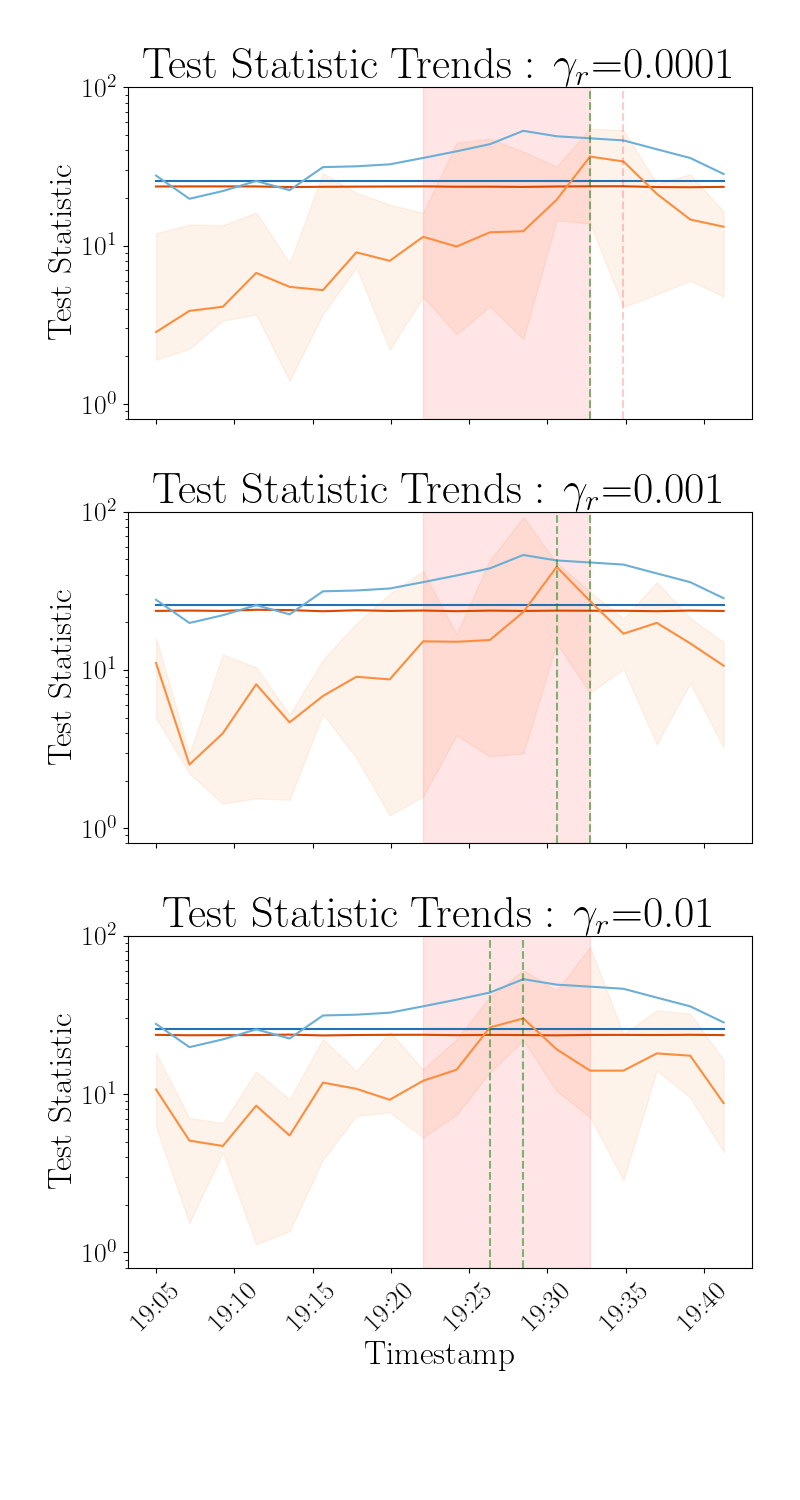}\label{fig:P1_ts_gr}}
        
        \vspace{-3.0em}
        \includegraphics[width=0.9\textwidth]{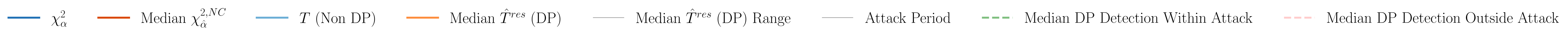}
        \vspace{3.0em}
    
  \caption{HAI Dataset: Test statistic trends for varying DP parameter values}\label{fig:P1_ts}
\vspace{-5mm}
\end{figure}

\begin{figure}[!htb]
    \centering
        \subfigure[$\epsilon_{cov}$]{\includegraphics[width=0.245\textwidth,keepaspectratio]{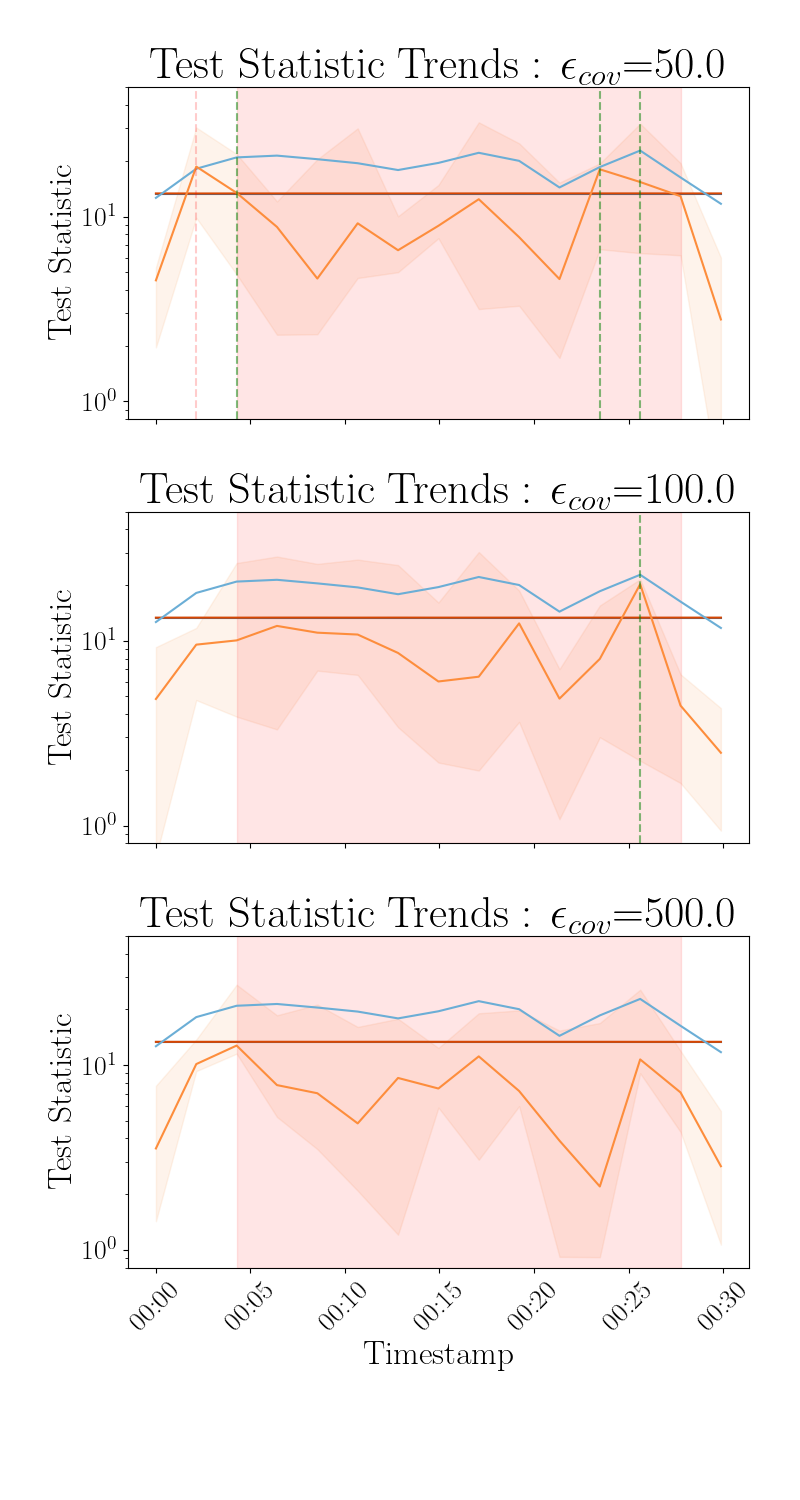}\label{fig:ps_ts_ec}}
        \hfill
        \subfigure[$\epsilon_{r}$]{\includegraphics[width=0.245\textwidth,keepaspectratio]{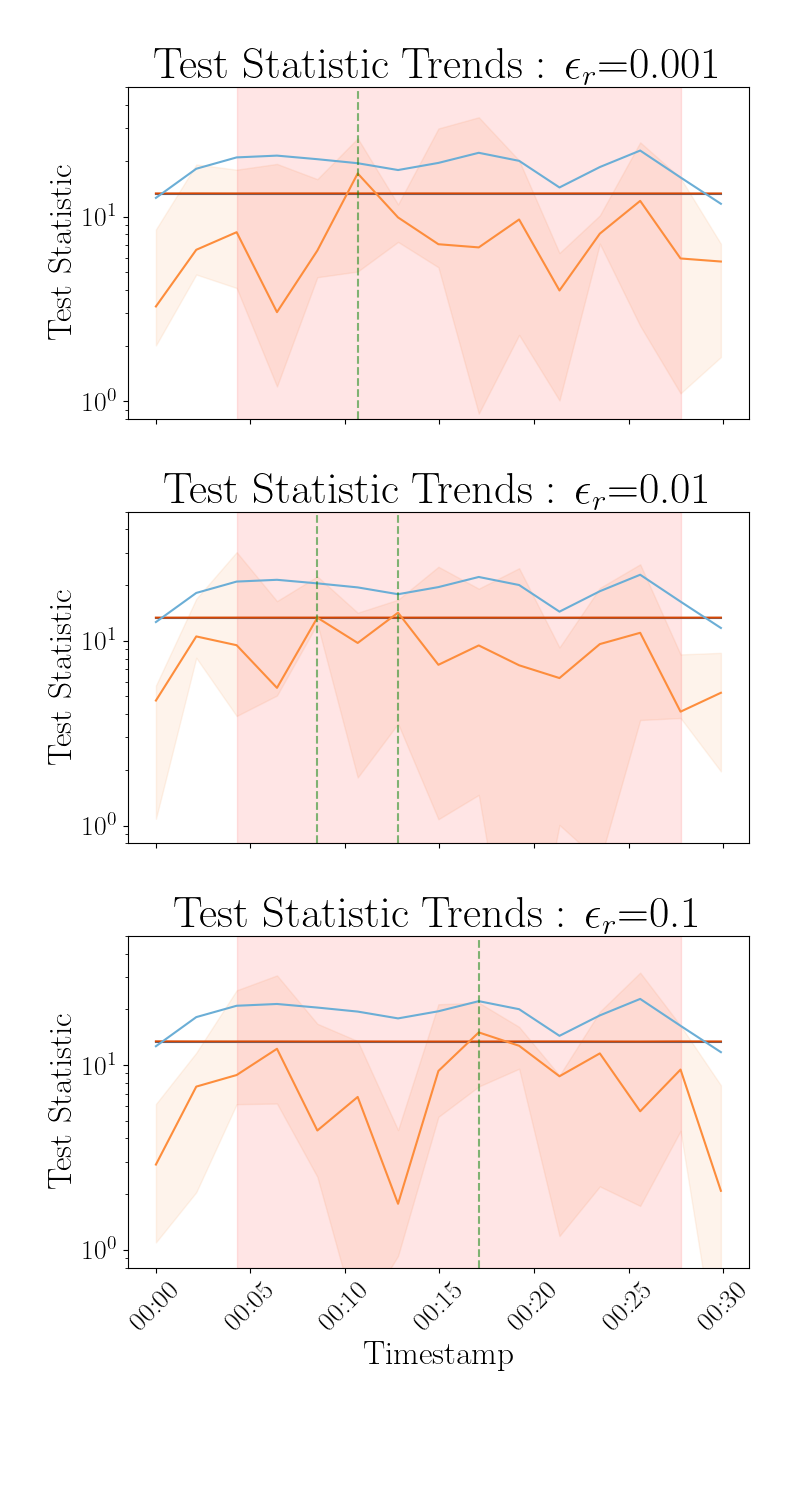}\label{fig:ps_ts_er}}
        \hfill
        \subfigure[$\gamma_{cov}$]{\includegraphics[width=0.245\textwidth,keepaspectratio]{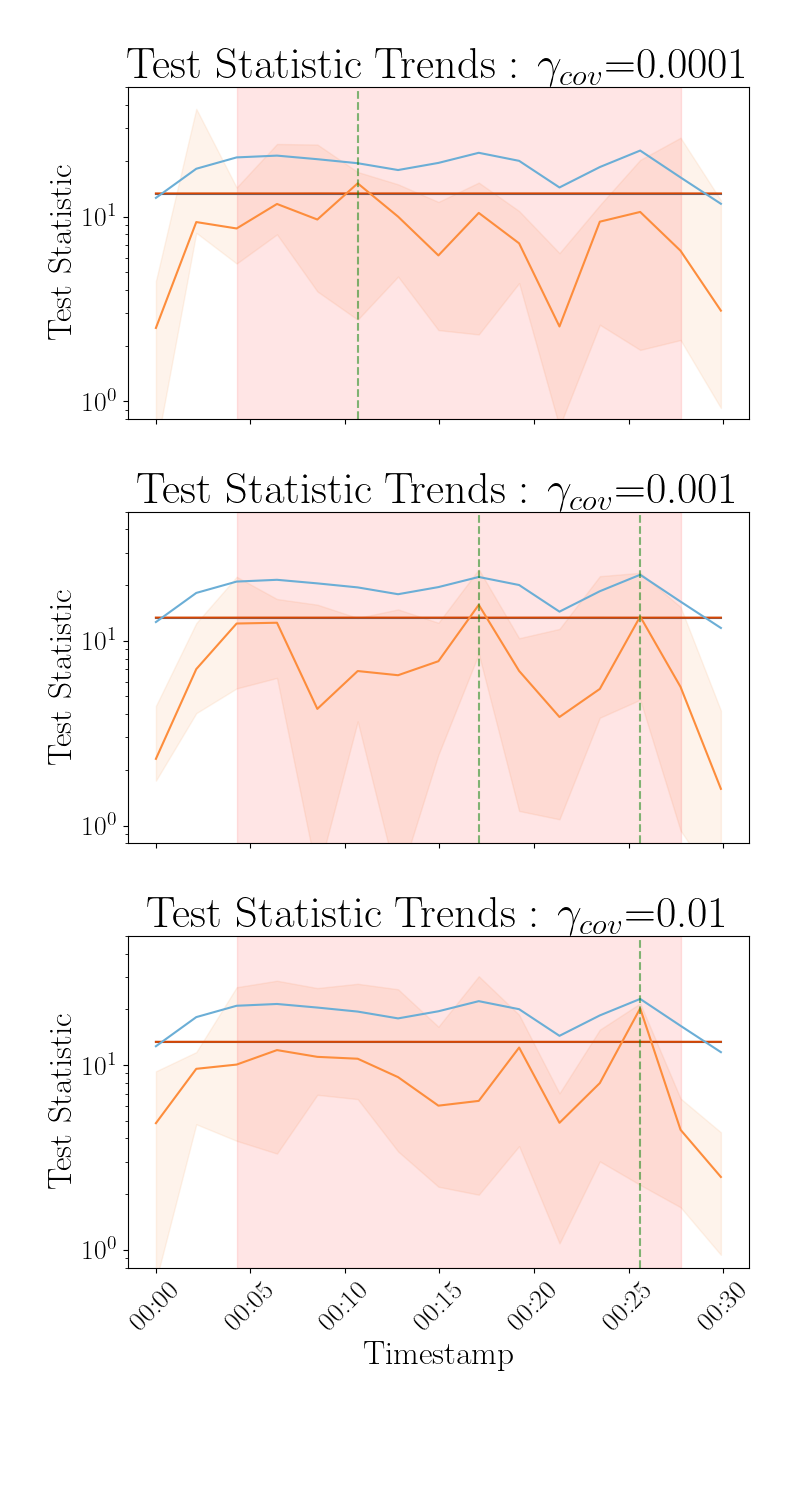}\label{fig:ps_ts_gc}}
        \hfill
        \subfigure[$\gamma_{r}$]{\includegraphics[width=0.245\textwidth,keepaspectratio]{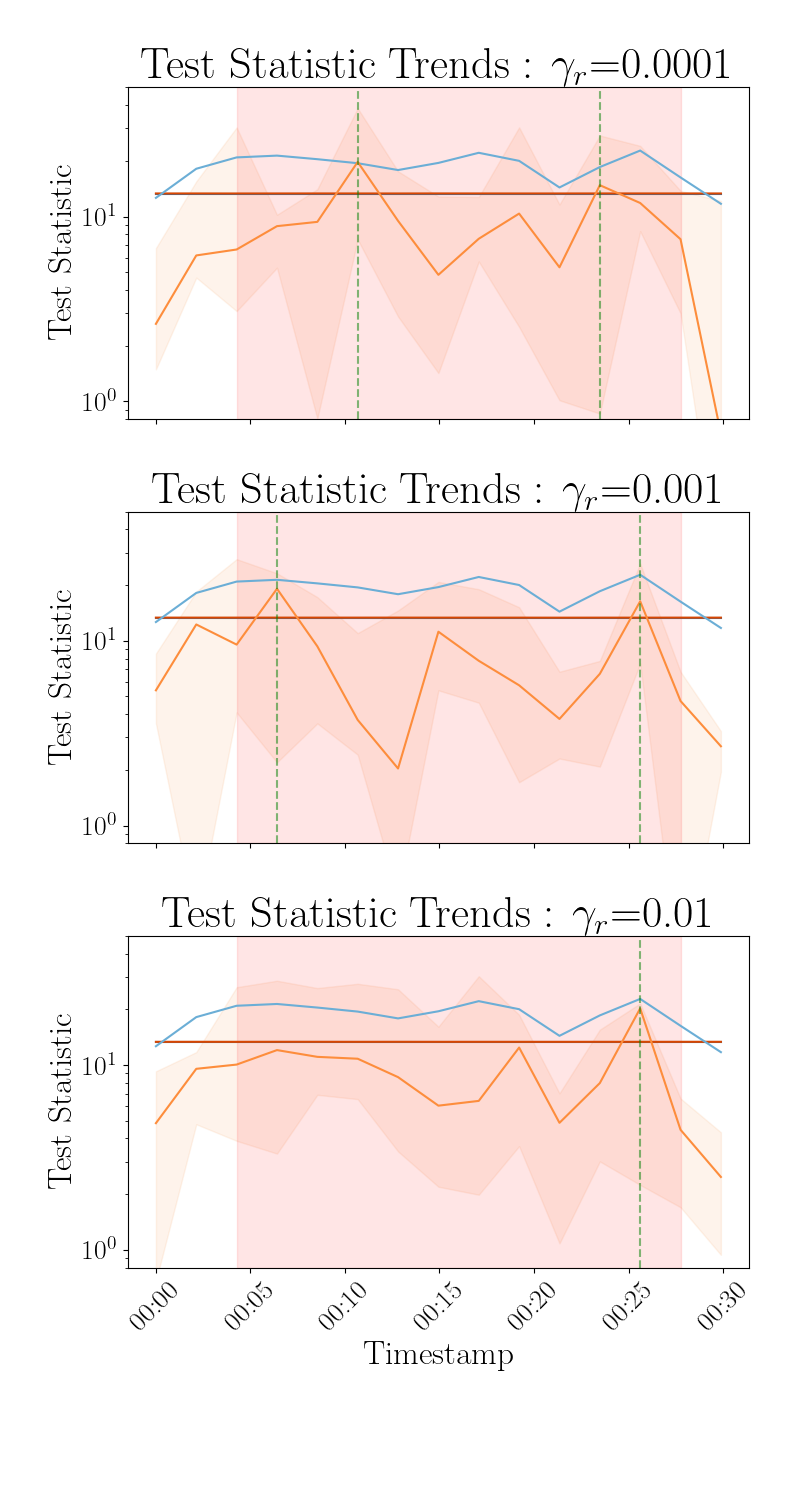}\label{fig:ps_ts_gr}}
        
        \vspace{-3.0em}
        \includegraphics[width=0.9\textwidth]{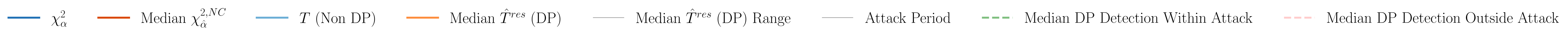}
        \vspace{3.0em}
        
  \caption{ORNL-PS Dataset: Test statistic trends for varying DP parameter values}\label{fig:ps_ts}
\end{figure}

\begin{table*}[ht]
\centering
\caption{HAI Dataset: Alignment of DP and Non-DP Detection Within Attack Window.}\label{tab:align_hai}
\renewcommand{\arraystretch}{1.05}
\setlength{\tabcolsep}{5pt}
\resizebox{\textwidth}{!}{%
\begin{tabular}{|c|c|c|c|c|c|c|c|c|c|c|c|c|}
\hline
\multirow{2}{*}{\makecell{\textbf{DP} \textbf{Params}}}
& \multicolumn{3}{c|}{\makecell{\textbf{DP \& Non-DP} \\ \textbf{Detection}}} 
& \multicolumn{3}{c|}{\makecell{\textbf{Only Non-DP} \\ \textbf{ Detection}}} 
& \multicolumn{3}{c|}{\makecell{\textbf{DP Alignment} \\ \textbf{Rate}}} 
& \multicolumn{3}{c|}{\makecell{\textbf{Mean $\hat{\alpha}$} \textbf{(Variance)} \\ in $1e-4$ ($1e-10$)} } \\
\cline{2-13}
& \textbf{200s} & \textbf{400s} & \textbf{600s} & \textbf{200s} & \textbf{400s} & \textbf{600s} & \textbf{200s} & \textbf{400s} & \textbf{600s} & \textbf{200s} & \textbf{400s} & \textbf{600s}\\ 
\hline
$\epsilon_{cov}=100$ & 11 & 26 & 44 & 39 & 24 & 6 & 0.22 & 0.52 & 0.88 & 2.56 (4.55) & 2.56 (4.55) & 2.56 (4.55)\\
$\epsilon_{r}=1e-3$ & 13 & 29 & 46 & 37 & 21 & 4 & 0.26 & 0.58 & 0.92 & 2.56 (4.5) & 2.56 (4.5) & 2.56 (4.5)\\
$\gamma_{cov}=1e-2$ & 8 & 25 & 44 & 42 & 25 & 6 & 0.16 & 0.5 & 0.88 & 2.57 (4.6) & 2.57 (4.6) & 2.57 (4.6)\\
$\gamma_{r}=1e-2$ & 11 & 27 & 45 & 39 & 23 & 5 & 0.22 & 0.54 & 0.9 & 2.56 (4.39) & 2.56 (4.39) & 2.56 (4.39)\\
\hline
\end{tabular}
}
\end{table*}

\begin{table*}[ht]
\centering
\caption{ORNL-PS Dataset: Alignment of DP and Non-DP Detection Within Attack Window}\label{tab:align_ornl}
\renewcommand{\arraystretch}{1.05}
\setlength{\tabcolsep}{5pt}
\resizebox{\textwidth}{!}{%
\begin{tabular}{|c|c|c|c|c|c|c|c|c|c|c|c|c|}
\hline
\multirow{2}{*}{\makecell{\textbf{DP} \textbf{Params}}}
& \multicolumn{3}{c|}{\makecell{\textbf{DP \& Non-DP} \\ \textbf{Detection}}} 
& \multicolumn{3}{c|}{\makecell{\textbf{Only Non-DP} \\ \textbf{ Detection}}} 
& \multicolumn{3}{c|}{\makecell{\textbf{DP Alignment} \\ \textbf{Rate}}} 
& \multicolumn{3}{c|}{\makecell{\textbf{Mean $\hat{\alpha}$} \textbf{(Variance)} \\ in $1e-4$ ($1e-6$)} } \\
\cline{2-13}
& \textbf{200s} & \textbf{400s} & \textbf{600s} & \textbf{200s} & \textbf{400s} & \textbf{600s} & \textbf{200s} & \textbf{400s} & \textbf{600s} & \textbf{200s} & \textbf{400s} & \textbf{600s}\\ 
\hline
$\epsilon_{cov}=100$ & 28 & 45 & 48 & 22 & 5 & 2 & 0.56 & 0.90 & 0.96 & 1.014 (1.0) & 1.014 (1.0) & 1.014 (1.0)\\
$\epsilon_{r}=1e-3$ & 28 & 45 & 47 & 22 & 5 & 3 & 0.56 & 0.90 & 0.94 & 0.976 (5.8) & 0.976 (5.8) & 0.976 (5.8)\\
$\gamma_{cov}=1e-2$ & 25 & 42 & 46 & 25 & 8 & 4 & 0.5 & 0.84 & 0.92 & 1.014 (1.0) & 1.014 (1.0) & 1.014 (1.0)\\
$\gamma_{r}=1e-2$ & 27 & 44 & 47 & 23 & 6 & 3 & 0.54 & 0.88 & 0.94 & 1.014 (1.0) & 1.014 (1.0) & 1.014 (1.0)\\
\hline
\end{tabular}
}
\end{table*}

\begin{table}[!htb]
\centering
\caption{DP False Alarm Rates Outside Attack Window}\label{tab:dp_false}
\renewcommand{\arraystretch}{1.05}  
\setlength{\tabcolsep}{4pt}        
\begin{tabular}{|c|c|c|c|c|}
\hline
\makecell{\textbf{Dataset}} &
$\epsilon_{cov}$=100 & $\epsilon_{r}$=1e-3 & $\gamma_{cov}$=1e-2 & $\gamma_{r}$=1e-2
 \\
\hline
HAI & 0.0767 & 0.0833 & 0.0783 & 0.08\\
ORNL-PS & 0.04 & 0.068 & 0.053 & 0.047\\ 
\hline
\end{tabular}
\vspace{-3mm}
\end{table}

\noindent
\textbf{DP parameter choice}: In our experiments $\frac{\Delta_{l}}{\epsilon_{cov}}$ denotes the scale value for the Laplacian distribution used for covariance disclosures. For all our experiments we utilize $\sigma = \frac{\Delta_r}{\epsilon_r}\sqrt{2.ln\Big(\frac{1.25}{\gamma_r}\Big)}$. For the HAI and the ORNL-PS datasets, we used $\Delta_r=50$, and $\Delta_l=0.1$. These values were chosen based on rigorous empirical analysis of the maximum 2-norm residual values observed as well as the maximum eigenvalue square roots observed in the covariance matrices. 

\noindent
\textbf{Visualization strategy}: We present graphs depicting trends that include a \emph{min-max} band comprising the minimum and maximum values of the desired quantity observed over five consecutive runs. The \emph{min-max} encapsulate the median values observed across five independent runs for each experiment. To accommodate the non-linear nature of the model and complexities in the HAI data, we consider a 40 minute window pertaining to the \texttt{ap\_05} attack scenario. In order to account for the nonlinear nature of the detection model, we designate the first 800 secs from the start \texttt{ap\_05} as the attack window for the HAI dataset. Since the attack duration in the ORNL-PS datasets exceeds 800 seconds, we restrict ourselves to results for the first 57 minutes from the beginning of the dataset. Red vertical lines depict false alarms with respect to the median obtained from five independent runs. Green vertical lines reflect the correct detection of an attack.

\subsection{Analyzing DP Covariance Disclosures}
We begin by analyzing the effects of differentially private disclosures of covariance matrices. Our analysis focuses on both HAI and ORNL-PS datasets specifically examining impacts of $\epsilon_{cov}$ and $\gamma_{cov}$ on the P-value and the test statistic in comparison to the non-DP scenarios.
\subsubsection{Effect of Privacy Budget}
We begin by analyzing the P-values of both datasets under varying covariance privacy budget values. Figures \ref{fig:P1_pv_ec}, \ref{fig:ps_pv_ec} reflect the P-value trends pertaining to HAI and ORNL-PS datasets respectively. The corresponding test statistic trends are depicted in Figures \ref{fig:P1_ts_ec} and \ref{fig:ps_ts_ec} respectively. 

We note that increasing values of $\epsilon_{cov}$ correspond to lower DP noise and a higher privacy budget as well. However, Figures \ref{fig:P1_pv_ec}, \ref{fig:ps_pv_ec}, \ref{fig:P1_ts_ec} and \ref{fig:ps_ts_ec} also indicate that with higher privacy budget, there is a much higher variation observed in both P-value and test statistic trends. In other words, with a higher privacy budget, we can observe much less variation in detection quality. Additionally, the consistency of alarms is higher in median terms during the attack period although this trend also results in some false alarms from the median as well. In Figure \ref{fig:ps_pv_ec}, we observe that for $\epsilon_{cov}=500$ the median does not breach the P-value threshold, although the trends make it clear that the attack detection is still robust with far lesser variations observed. 

\subsubsection{Effect of Failure Probability}
We turn our attention to the effect of failure probability observed when disclosing covariance values. Figures \ref{fig:P1_pv_gc}, \ref{fig:ps_pv_gc} represent trends in the P-value for HAI and ORNL-PS datasets, while Figures \ref{fig:P1_ts_gc}, \ref{fig:ps_ts_gc} represent trends in test statistic. $\gamma_{cov}$ conventionally represents the DP failure probability for covariance disclosures. Ultimately, a higher $\gamma_{cov}$ represents a higher likelihood of DP being ineffective at hiding. From Figures \ref{fig:P1_pv_gc}, \ref{fig:ps_pv_gc}, \ref{fig:P1_ts_gc} and \ref{fig:ps_ts_gc}, we observe that with increasing values of $\gamma_{cov}$, we again see more variance and in P-value and test statistic trends. However, we can also observe that the median detection quality results in consistent alarms during the attack window albeit with a higher variance. In summary, PRECISE is capable of providing attack detection with realistic privacy expectations even in cases of high $\gamma_{cov}$ values, indicating the usefulness of our approach. 

\subsection{Analysis for DP Residual Disclosures}
We now analyze the impact of differentially private residual disclosures in terms of the privacy budget $\epsilon_r$ and failure probability $\gamma_r$. 
\subsubsection{Effect of Privacy Budget}
We plot the P-value trends for varying values of $\epsilon_{r}$ for both datasets in Figures \ref{fig:P1_pv_er} and \ref{fig:ps_pv_er} respectively. Similarly, we present test statistic trends for both datasets in \ref{fig:P1_ts_er} and \ref{fig:ps_ts_er} respectively as well. In general, we see that the spread of values as indicated by the min-max spread increases with decreasing values of privacy budget $\epsilon_{r}$. This results in a slight increase in median alarm consistency for higher values of $\epsilon_{r}$ as well. Overall, even with a high value of privacy budget, we achieve significant privacy guarantee.

\subsubsection{Effect of Failure Probability}
We present the performance of the DP detection framework in terms of the residual disclosure failure probability $\gamma_r$ in terms of the P-value and test statistics. For both datasets, we can see that the attack detection performance remains robust with steady consistency of the median alarm detection rates as well. 

\subsection{Analyzing Attack Detection Quality}
In this subsection, we analyze the performance of the combined DP attack detection framework considering the residual based DP disclosures in conjunction with the DP based covariance matrices. The combined analysis is meant to provide vital insights into the latency of attack detection at the regulator level with respect to the local utility. 

Therefore, we examine the ability of the DP based framework to detect the attack within three distinct intervals (200s, 400s and 600s) measured from the start of the attack. The results of these experiments are obtained from five independent runs carried out for each DP parameter combination listed in Tables \ref{tab:align_hai} and \ref{tab:align_ornl}. We are primarily interested in tracking the alignment of DP and Non-DP detection at the discrete interval values of 200s, 400s and 600s. The alignment problem can be thought of as run instances wherein both DP and Non-DP frameworks successfully detected an attack. Therefore, we present results in terms of alignment of DP and Non-DP Detection, Non-DP only detection, DP alignment rates as well as the mean and variance of the $\hat{\alpha}$. 

In Tables \ref{tab:align_hai} and \ref{tab:align_ornl}  we see a consistent improvement in the alignment rate with increase in duration from the beginning of the attack. The alignment rate measured as a fraction of runs where an attack was detected within the attack window consistently improves from around 0.5 to 0.95. This is powered by the rising number of DP and Non-DP detection instances that is naturally accompanied by falling Non-DP only detection. For both datasets, we see that the mean $\hat{\alpha}$ stays relatively stable at $2.56e-4$ and $1.01e-4$ respectively. The associated variance experiences minor volatility but overall retains stability around $4.55e-10$ and $1e-10$ for HAI and ORNL-PS datasets respectively.

In Table \ref{tab:dp_false}, we present the false alarm rates of the DP based detection framework for both datasets captured during normal operations. We can see that for all the considered combinations of DP parameters, the false alarm rates stay consistently low. For the HAI dataset, the stability of false alarm rates hover around $8\%$, while for ORNL-PS dataset, this value exhibits slightly more volatility, ranging from $4\%$ to around $6.8\%$. Collectively, Tables \ref{tab:align_hai}, \ref{tab:align_ornl} and \ref{tab:dp_false} demonstrate the robustness of the attack detection quality with respect to the misclassification and alignment rates of the DP mechanism.

\section{Conclusion}
In this paper, we present PRECISE, framework geared towards regulatory compliance for detecting data-driven attacks in industrial control systems for critical infrastructure networks. Our proposed method leverages statistical tests on residuals arising out of state space modeling at the utility stakeholder level to raise attack alarms. We focus on cases wherein utilities are interested in convincing regulatory bodies regarding the veracity of their respective alarms by disclosing differential privacy induced covariance matrices and residual values. As a result, PRECISE revolves around a two phase privacy scheme that sequentially perturbs covariance using Laplacian noise followed by a Gaussian differential privacy scheme for residuals. We derive strong privacy guarantees pertaining to the test of residuals in addition to providing tight bounds on the misclassification rates of alarms as well as equivalent levels of significance. We specifically explore two significant modalities of implementation concerning critical region and P-value based compliance schemes. Additionally, we theoretically characterize the privacy implications of each of the modalities. Using real-world ICS datasets, we characterize the performance of our algorithm with respect to varying privacy parameters under diverse attack scenarios. The experimental results demonstrate that PRECISE is capable of matching the performance of the non-DP versions in almost all cases while preserving the privacy of utility stakeholders.



\bibliographystyle{ACM-Reference-Format}
\bibliography{main}

\appendix
\section{Proofs of Lemmas and Theorems}
\subsection{Proof of Lemma \ref{thm1}}\label{subsec:pf_thm1}
\begin{proof}
We begin by stating Equation \eqref{eq:pce1} which results from the orthonormal factorization of the real and perturbed covariance matrices $S_t$ and $\hat{S}_t$ respectively as defined in Section \ref{subsec:htad}. 
\begin{gather}
    |\tau^T_t\tau_t - (\hat{\tau}^{cov})^T_t\hat{\tau}^{cov}_t| = |r^T_t (\Lambda_t-\hat{\Lambda}_t)r_t| \label{eq:pce1}
\end{gather}
We know that Equation \eqref{eq:lm11} holds when $\hat{\lambda}^{PCA},\lambda^{PCA}$ denote the diagonal matrices of eigenvalues corresponding to the top $p\leq d$ principal components with and without DP respectively
\begin{gather}\label{eq:lm11}
|r^T_t (\Lambda-\hat{\Lambda})r_t| = \Big|\sum\limits^{p}_{i=1} (r^i_t)^2 (\lambda_i - \hat{\lambda}_i) \Big| \leq E_l \Big|\sum\limits^{p}_{i=1} (r^i_t)^2 \Big|
\end{gather}
We know that Equation \eqref{eq:lm12} holds with a probability of at least $1-\gamma_{cov}$
\begin{gather}\label{eq:lm12}
    E_l\Big|\sum\limits^{p}_{i=1} (r^i_t)^2 \Big| \leq \Big|\sum\limits^{p}_{i=1} (r^i_t)^2 \Big|\Big[\frac{\Delta_l}{\epsilon_{cov}}log\Big(\frac{d}{\gamma_{cov}}\Big)\Big]
\end{gather}
By combining Equations \eqref{eq:lm11}, \eqref{eq:lm12} with Equation \eqref{eq:pce2}, we get Equation \eqref{eq:lm13} which holds with probability of at least $1-\gamma_{cov}$ completing the proof.
\begin{dmath}\label{eq:lm13}
|\hat{T}^{cov}_{\chi^2,t} - T_{\chi^2,t}| \leq \Big|\sum\limits^{p}_{i=1} (r^i_t)^2 \Big|\Big[\frac{\Delta_l}{\epsilon_{cov}}log\Big(\frac{d}{\gamma_{cov}}\Big)\Big]
\end{dmath}
\end{proof}
\vspace{-5mm}
\subsection{Proof of Lemma \ref{lem1}}\label{subsec:pf_lem1}
\begin{proof}
As a result of Definition \ref{defn3}, we obtain an $(\epsilon,\gamma_r)$-DP mechanism such that Equation \eqref{eq:res1} applies.
\begin{gather}\label{eq:res1}
\mathbb{P}\Big(\Big|\sum\limits_{i=1}^{p} e_{i,t} \Big| \geq \theta_r \Big) \leq \gamma_r,\text{ where, }\theta_r = \frac{\sigma^2\epsilon}{\Delta} - \frac{p\Delta}{2}
\end{gather}

Since $\sum\limits_{i=1}^{p} |e_{i,t}| \geq|\sum\limits_{i=1}^{p} e_{i,t}|$, we can also derive the following relations
\begin{gather} \label{eq:pres3}
\mathbb{P}\Big(E_r \leq \frac{\theta_r}{p}\Big) \geq 1-\gamma_r, \text{ where, } E_r = \underset{0\leq i\leq p}{max}|e_{i,t}|
\end{gather}
Equation \eqref{eq:pres3} ensures that the probability of the maximum absolute value of Gaussian noise being less than $\theta_r/p$ is at least $1-\gamma_r$. Using Equation \eqref{eq:pres3} we can bound $E^2_r$ as well.
\begin{gather}
\mathbb{P}\Big(0 \leq E^2_r \leq \frac{\theta_r^2}{p^2}\Big) \geq 1-\gamma_r
\end{gather}
Since each of the elements of $e_t$ are independently and identically distributed Equation \eqref{eq:pres4} holds
\begin{gather}\label{eq:pres4}
\mathbb{P}\Big(0 \leq \sum\limits_{i=1}^{p} e^2_{i,t} \leq p.E^2_r \leq \frac{\theta_r^2}{p}\Big) \geq (1-\gamma_r)^p
\end{gather}
Setting $||e_t||^2_2 = \sum\limits_{i=1}^{p} e^2_{i,t}$ in Equation \eqref{eq:pres4} completes the proof.
\end{proof}
\vspace{-5mm}
\subsection{Proof of Lemma \ref{lem2}}\label{subsec:pf_lem2}
\begin{proof}
We know that, $\hat{T}^{res}_{\chi^2,t}  = \sum\limits_{i=1}^{p}(e_{i,t} + \tau_{i,t})^2$. From GDP conditions, $(e_{i,t} + \tau_{i,t}) \sim N(\tau_{i,t},\sigma^2)$. As a result, $[(e_{i,t} + \tau_{i,t})^2/\sigma^2] \sim \chi^2(1,\tau^2_{i,t}/\sigma^2)$ which implies that the variance scaled perturbed test statistic $\hat{T}^{res}_{\chi^2,t}/\sigma^2\sim \chi^{2}(p,||\tau_t||^2_2/\sigma^2)$. 
\end{proof}
\vspace{-5mm}

\subsection{Proof of Theorem \ref{thm2}}\label{subsec:pf_thm2}
\begin{proof}
We reformulate each element of $\hat{T}^{res}_{\chi^2,t} - T_{\chi^2,t}$ as 
\begin{equation}
\begin{aligned}\label{eq:thm21}
\hat{T}^{res}_{\chi^2,t} - T_{\chi^2,t} & = \sum\limits_{i=1}^{p}e_{i,t}(e_{i,t} + 2\tau_{i,t})\\ &= 2\sum\limits_{i=1}^{p}(e_{i,t} \tau_{i,t}) + \sum\limits_{i=1}^{p} (e_{i,t})^2
\end{aligned}
\end{equation}
Using Lemma 1, we can assert that $e_{i,t}\leq E_r \leq \frac{\theta_r}{p}$ and $\sum\limits_{i=1}^{p} (e_{i,t})^2 \leq \frac{\theta_r^2}{p}$ with a probability of at least $(1-\gamma_r)^p$. Therefore, combining Equation \eqref{eq:thm21} with Lemma 1, we get
\begin{gather}
2\sum\limits_{i=1}^{p}(e_{i,t} \tau_{i,t}) + \sum\limits_{i=1}^{p} (e_{i,t})^2 \leq \frac{\theta_r}{p}\Big(\theta_r+2\sum\limits_{i=1}^{p}\tau_{i,t}\Big)\\
2\sum\limits_{i=1}^{p}(e_{i,t} \tau_{i,t}) + \sum\limits_{i=1}^{p} (e_{i,t})^2\geq \frac{\theta_r}{p}\Big(\theta_r-2\sum\limits_{i=1}^{p}\tau_{i,t}\Big)
\end{gather}
Using Lemma 2, for $U' = \frac{\theta_r}{p}\Big(\theta_r+2\sum\limits_{i=1}^{p}\tau_{i,t}\Big)$, $L' = \frac{\theta_r}{p}\Big(\theta_r-2\sum\limits_{i=1}^{p}\tau_{i,t}\Big)$ we can also assert that
\begin{gather}
\mathbb{P}\Big[ \frac{1}{\sigma^2}(\hat{T}^{res}_{\chi^2,t} - T_{\chi^2,t}) \leq \frac{U'}{\sigma^2} \Big| ||e_t||^2_2 \leq \frac{\theta_r^2}{p} \Big] = F^{\tau_t}_{\chi^2,p}(\frac{U'}{\sigma^2})\\
\mathbb{P}\Big[ \frac{1}{\sigma^2}(\hat{T}^{res}_{\chi^2,t} - T_{\chi^2,t}) \geq \frac{L'}{\sigma^2} \Big| ||e_t||^2_2 \leq \frac{\theta_r^2}{p} \Big] = 1 - F^{\tau_t}_{\chi^2,p}(\frac{L'}{\sigma^2})
\end{gather}
Computing joint probability using Lemma 1 and setting $U=U'/\sigma^2$ and $L=L'/\sigma^2$ completes the proof 
\begin{align*}
{\mathbb{P}\Big[ L \leq (\hat{T}^{res}_{\chi^2,t} - T_{\chi^2,t}) \leq U \Big|||e_t||^2_2 \leq \frac{\theta_r^2}{p}\Big]}\\ \geq (F^{\tau_t}_{\chi^2,p}(U) - F^{\tau_t}_{\chi^2,p}(L)) (1-\gamma_{r})^p
\end{align*}
\end{proof}

\subsection{Proof of Lemma \ref{lem3}}\label{subsec:pf_lem3}
\begin{proof}
We know that
\begin{gather}
\mathbb{P}\Big[ \hat{T}^{res}_{\chi^2,t} > \phi \Big] = 1-F^{\tau}_{\chi^2,p}(\phi)
\end{gather}
The CDF of $\chi^2(k,\mu)$ can be stated as $F^{\tau_t}_{\chi^2,p}(\phi) = 1 - Q_{\frac{k}{2}}(\sqrt{\mu},\sqrt{\phi})$ where $Q_{a}(b,c)$ is the generalized Marcum Q-function. The generalized Marcum Q-function is strictly increasing in $\mu$ \cite{sun2010monotonicity}. As a result,
\begin{gather}
\underset{\tau}{max}(1-F^{\tau}_{\chi^2,p}(\phi)) = \underset{\tau}{max}\text{ }Q_{\frac{p}{2}}\Big(\frac{||\tau||_2}{\sigma},\sqrt{\phi}\Big)
\end{gather}
As a result, given $\tau_{min} = \underset{\tau}{argmin}(||\tau||_2)$ and $\tau_{max} = \underset{\tau}{argmax}(||\tau||_2)$

\begin{gather}
\tau_{max} = \underset{\tau}{argmax}\text{ }(1-F^{\tau}_{\chi^2,p}(\phi))
\end{gather}
\end{proof}

\subsection{Proof of Theorem \ref{thm3}}\label{subsec:pf_thm3}
\begin{proof}
We consider two cases\\
\textit{Case1}: When $|\hat{T}^{cov}_{\chi^2,t} - T_{\chi^2,t}|\geq R_t\theta_l$, we know that
\begin{equation} \label{eq:thm31}
\begin{aligned}
\mathbb{P}\Big[ \hat{T}^{res}_{\chi^2,t} > \sigma^2\chi^{2,NC}_{\hat{\alpha}},\ 
|\hat{T}^{cov}_{\chi^2,t} - T_{\chi^2,t}|\geq R_t\theta_l \Big] = \\ 
\mathbb{P}\Big[|\hat{T}^{cov}_{\chi^2,t} - T_{\chi^2,t}|\geq R_t\theta_l\Big]\Big[ 1- F^{\hat{\tau}^{cov}_t}_{\chi^2,p}
\Big(\sigma^2\chi^{2,NC}_{\hat{\alpha}}\Big)\Big]
\end{aligned}
\end{equation}
Using Lemma \ref{thm1}, we can state that 
\begin{gather}
\mathbb{P}\Big[|\hat{T}^{cov}_{\chi^2,t} - T_{\chi^2,t}|\geq R_t\theta_l\Big] = \mathbb{P}\Big[|r^T_t(\lambda_{t,i}-\hat{\lambda}_{t,i})r_t| \geq R_t\theta_l\Big]
\end{gather}
We note that using Lemma \ref{thm1}, we can know that since $|\hat{T}^{cov}_{\chi^2,t} - T_{\chi^2,t}| = |r^T_t(\Lambda_t-\hat{\Lambda}_t)r_t|$, we can state that using $r_{max} = \underset{r}{argmax} ||r||$
\begin{gather}
|r^T_t(\Lambda_t-\hat{\Lambda}_t)r_t| \leq |\sum\limits_{i=1}^{p}(\lambda_t-\hat{\lambda}_{t,i}) r^2_{t,i}|\\
|\sum\limits_{i=1}^{p}(\lambda_t-\hat{\lambda}_{t,i}) r^2_{t,i}| \leq \sum\limits_{i=1}^{p}|(\lambda_t-\hat{\lambda}_{t,i})| ||r_{max}||^2_2 
\end{gather}
As a result, we know that $(\lambda_t-\hat{\lambda}_{t,i}) \sim Lap(0,\frac{\Delta_l}{\epsilon_{cov}})$, which implies that $r^2_{max} |(\lambda_t-\hat{\lambda}_{t,i})| \sim Exp(\frac{\epsilon_{cov}}{\Delta_lr^2_{max}})$ leading to
\begin{gather}
\Big[\sum\limits_{p}|(\lambda_{t,i}-\hat{\lambda}_{t,i})|\Big]||r_{max}||^2_2  \sim Gamma(p,\frac{\epsilon_{cov}}{\Delta_lr^2_{max}})
\end{gather}
Since, we know that $|r^T_t(\Lambda_t-\hat{\Lambda}_t)r_t| =|\hat{T}^{cov}_{\chi^2,t} - T_{\chi^2,t}|$, we can state that,
$\Big[\sum\limits_{p}|(\lambda_{t,i}-\hat{\lambda}_{t,i})|\Big]||r_{max}||^2_2 \geq |r^T_t(\Lambda_t-\hat{\Lambda}_t)r_t|$.
Probabilistically, this leads us to,
\begin{equation}
    \begin{aligned}
        \mathbb{P}\Big[\Big(\sum\limits_{p}|(\lambda_{t,i}-\hat{\lambda}_{t,i})|\Big)||r_{max}||^2_2& \geq|\hat{T}^{cov}_{\chi^2,t} - T_{\chi^2,t}| \geq R_t\theta_l\Big]\\& \leq 1-F^{gamma}_{r_{max},\epsilon_{cov},\Delta_l}(R_t\theta_l)
    \end{aligned}
\end{equation}

Therefore we get Equation \eqref{eq:thm31}, where $F^{gamma}_{r_{max},\epsilon_{cov},\Delta_l} (.)$ represents the CDF of $Gamma(p,\frac{\epsilon_{cov}}{\Delta_lr^2_{max}})$
\begin{equation} \label{eq:thm31a}
\begin{aligned}
\mathbb{P}\Big[ \hat{T}^{res}_{\chi^2,t} > \sigma^2\chi^{2,NC}_{\hat{\alpha}},\ 
|\hat{T}^{cov}_{\chi^2,t} - T_{\chi^2,t}|\geq R_t\theta_l \Big] \leq \\ 
\Big[1-F^{gamma}_{r_{max},\epsilon_{cov},\Delta_l}(R_t\theta_l)\Big]\Big[ 1- F^{\hat{\tau}^{cov}_t}_{\chi^2,p}
\Big(\sigma^2\chi^{2,NC}_{\hat{\alpha}}\Big)\Big]
\end{aligned}
\end{equation}

\textit{Case2}: For the case when $|\hat{T}^{cov}_{\chi^2,t} - T|\leq R_t\theta_l$
Using Lemma \ref{thm1}, we can state that 
\begin{gather}
\mathbb{P}\Big[|\hat{T}^{cov}_{\chi^2,t} - T_{\chi^2,t}|\leq R_t\theta_l\Big] = \mathbb{P}\Big[|r^T_t(\lambda_{t,i}-\hat{\lambda}_{t,i})r_t| \leq R_t\theta_l\Big]
\end{gather}
We note that $|r^T_t(\lambda_{t,i}-\hat{\lambda}_{t,i})r_t| \leq R_t\theta_l = r^T_tdiag(\theta_l)r_t$ holds for all $r$. This implies that $|(\lambda_{t,i}-\hat{\lambda}_{t,i})| \leq \theta_l$, and as a result, 
\begin{gather}
\mathbb{P}\Big[|r^T_t(\Lambda_t-\hat{\Lambda}_t)r_t| \leq R_t\theta_l\Big] = \mathbb{P}\Big[|\Lambda_t-\hat{\Lambda}_t| \leq \theta_l\Big]
\end{gather}
We also know that $|\Lambda_t-\hat{\Lambda}_t| \sim Exp(\frac{\epsilon_{cov}}{\Delta_l})$, Therefore $\mathbb{P}\Big[|\Lambda_t-\hat{\Lambda}_t| \leq \theta_l\Big] = [F^{ex}_{\Delta_l,\epsilon_{cov}}(\theta_l)]^p$, where $F^{ex}_{\Delta_l,\epsilon_{cov}}(.)$ is the CDF of the exponential distribution. Therefore, we can state that,
\begin{equation} \label{eq:thm31b}
\begin{aligned}
\mathbb{P}\Big[ \hat{T}^{res}_{\chi^2,t} > \sigma^2\chi^{2,NC}_{\hat{\alpha}},\ 
|\hat{T}^{cov}_{\chi^2,t} - T_{\chi^2,t}|\leq R_t\theta_l \Big]
= \\ 
\Big[F^{ex}_{\Delta_l,\epsilon_{cov}}(\theta_l)\Big]^p\Big[ 1- F^{\hat{\tau}^{cov}_t}_{\chi^2,p}
\Big(\sigma^2\chi^{2,NC}_{\hat{\alpha}}\Big)\Big]
\end{aligned}
\end{equation}
Using Lemma \ref{lem3}, we get
\begin{equation} \label{eq:thm31c}
\begin{aligned}
\mathbb{P}\Big[ \hat{T}^{res}_{\chi^2,t} > \sigma^2\chi^{2,NC}_{\hat{\alpha}},\ 
|\hat{T}^{cov}_{\chi^2,t} - T_{\chi^2,t}|\leq R_t\theta_l \Big]
\leq \\ 
\Big[F^{ex}_{\Delta_l,\epsilon_{cov}}(\theta_l)\Big]^p\Big[ 1- F^{\hat{\tau}^{cov}_{max,t}}_{\chi^2,p}\Big(\sigma^2\chi^{2,NC}_{\hat{\alpha}}\Big)\Big]
\end{aligned}
\end{equation}
\end{proof}


\subsection{Proof of Theorem \ref{thm4}}\label{subsec:pf_thm4}
\begin{proof}
We know that $\hat{\rho}_t=1$ is triggered when $\hat{T}^{res}_{\chi,t} >\sigma^2\chi^{2,NC}_{\hat{\alpha}}$, and $\rho_t=0$ when $T_{\chi^2,t} < \chi^{2}_{\alpha}$. Therefore,
\begin{gather}
\hat{T}^{res}_{\chi^2,t} - T_{\chi^2,t} >  \sigma^2\chi^{2,NC}_{\hat{\alpha}} - \chi^{2}_{\alpha}\\
\hat{T}^{res}_{\chi^2,t} > T_{\chi^2,t} + \sigma^2\chi^{2,NC}_{\hat{\alpha}} - \chi^{2}_{\alpha}
\end{gather}
As a result, we can state that
\begin{gather}
\mathbb{P}[\hat{\rho}_t=1|\rho_t=0,\hat{\tau}^{cov}_{\chi^2,t}] = \mathbb{P}[\hat{T}^{res}_{\chi^2,t} > T_{\chi^2,t} + \sigma^2\chi^{2,NC}_{\hat{\alpha}} - \chi^{2}_{\alpha}]\\
\mathbb{P}[\hat{\rho}_t=1|\rho_t=0,\hat{\tau}^{cov}_{\chi^2,t}] = 1- F^{\hat{\tau}^{cov}_t}_{\chi^2,p}\Big(T_{\chi^2,t} + \sigma^2\chi^{2,NC}_{\hat{\alpha}} - \chi^{2}_{\alpha}\Big)
\end{gather}

We can use the same technique for upper bounding as employed in Cases 1 and 2 in Theorem \ref{thm3} in Equations \eqref{eq:thm31} and \eqref{eq:thm31b} for the event $\hat{T}^{res}_{\chi^2,t} > T_{\chi^2,t} + \sigma^2\chi^{2,NC}_{\hat{\alpha}} - \chi^{2}_{\alpha}$. This leads us to Equations \eqref{eq:thm41} and \eqref{eq:thm41b}.
\begin{equation} \label{eq:thm41}
\begin{aligned}
\mathbb{P}\Big[ \hat{T}^{res}_{\chi^2,t} > T_{\chi^2,t} + \sigma^2\chi^{2,NC}_{\hat{\alpha}} - \chi^{2}_{\alpha}, &
|\hat{T}^{cov}_{\chi^2,t} - T_{\chi^2,t}|\geq R_t\theta_l \Big]\\ 
\leq \Big[1-F^{gamma}_{r_{max},\epsilon_{cov},\Delta_l}(R_t\theta_l)\Big]\\\Big[ 1- F^{\hat{\tau}^{cov}_t}_{\chi^2,p}
\Big(T_{\chi^2,t}& + \sigma^2\chi^{2,NC}_{\hat{\alpha}} - \chi^{2}_{\alpha}\Big)\Big]
\end{aligned}
\end{equation}
\begin{equation} \label{eq:thm41b}
\begin{aligned}
\mathbb{P}\Big[ \hat{T}^{res}_{\chi^2,t} > T_{\chi^2,t} + \sigma^2\chi^{2,NC}_{\hat{\alpha}} - \chi^{2}_{\alpha}, &
|\hat{T}^{cov}_{\chi^2,t} - T_{\chi^2,t}|\leq R_t\theta_l \Big]\\ 
\leq \Big[F^{ex}_{\Delta_l,\epsilon_{cov}}(\theta_l)\Big]^p\Big[ 1- F^{\hat{\tau}^{cov}_{max,t}}_{\chi^2,p}
\Big(T_{\chi^2,t}& + \sigma^2\chi^{2,NC}_{\hat{\alpha}} - \chi^{2}_{\alpha}\Big)\Big]
\end{aligned}
\end{equation}
Summing up Equations \eqref{eq:thm41} and \eqref{eq:thm41b} leads us to the following where $\hat{T}_t = T_{\chi^2,t} + \sigma^2\chi^{2,NC}_{\hat{\alpha}} - \chi^{2}_{\alpha}$
\begin{equation} \label{eq:thm41c}
\begin{aligned}
\mathbb{P}[\hat{\rho}_t=1|&\rho_t=0] 
\leq\\ 
&\Big[1-F^{gamma}_{r_{max},\epsilon_{cov},\Delta_l}(R_t\theta_l)\Big]\Big[ 1-F^{\hat{\tau}^{cov}_t}_{\chi^2,p}\Big(\hat{T}_t\Big)\Big]\\ 
&+ \Big[F^{ex}_{\Delta_l,\epsilon_{cov}}(\theta_l)\Big]^p\Big[ 1- F^{\hat{\tau}^{cov}_{max,t}}_{\chi^2,p}
\Big(\hat{T}_t\Big)\Big]
\end{aligned}
\end{equation}
\noindent
Similarly, $\hat{\rho}_t=0$ when $\hat{T}^{res}_{\chi,t} <\sigma^2\chi^{2,NC}_{\hat{\alpha}}$, and $\rho_t=1$ when $T_{\chi^2,t} > \chi^{2}_{\alpha}$. Therefore,
\begin{gather}
\hat{T}^{res}_{\chi^2,t} - T_{\chi^2,t} <  \sigma^2\chi^{2,NC}_{\hat{\alpha}} - \chi^{2}_{\alpha}\\
\hat{T}^{res}_{\chi^2,t} < T_{\chi^2,t} + \sigma^2\chi^{2,NC}_{\hat{\alpha}} - \chi^{2}_{\alpha}
\end{gather}
In a similar fashion as Case 1, we can obtain,
\begin{gather}
\mathbb{P}[\hat{\rho}_t=0|\rho_t=1] = \mathbb{P}[\hat{T}^{res}_{\chi^2,t} < \hat{T}_t] = F^{\hat{\tau}^{cov}_t}_{\chi^2,p}\Big(\hat{T}_t\Big)
\end{gather}

Therefore, we obtain Equations \eqref{eq:thm41d} and \eqref{eq:thm41e}
\begin{equation} \label{eq:thm41d}
\begin{aligned}
\mathbb{P}\Big[ \hat{T}^{res}_{\chi^2,t} < \hat{T}_t,\ 
|\hat{T}^{cov}_{\chi^2,t} - T_{\chi^2,t}|\leq R_t\theta_l \Big]
\leq \\ 
\Big[F^{ex}_{\Delta_l,\epsilon_{cov}}(\theta_l)\Big]^pF^{\hat{\tau}^{cov}_t}_{\chi^2,p}\Big(\hat{T}_t\Big)
\end{aligned}
\end{equation}
\begin{equation} \label{eq:thm41e}
\begin{aligned}
\mathbb{P}\Big[ \hat{T}^{res}_{\chi^2,t} < \hat{T}_t,\ 
|\hat{T}^{cov}_{\chi^2,t} - T_{\chi^2,t}|\geq R_t\theta_l \Big]
\leq \\ 
\Big[1-F^{gamma}_{r_{max},\epsilon_{cov},\Delta_l}(R_t\theta_l)\Big]F^{\hat{\tau}^{cov}_t}_{\chi^2,p}\Big(\hat{T}_t\Big)
\end{aligned}
\end{equation}
Combining both equations leads us to
\begin{equation}
\begin{aligned}
&\mathbb{P}[\hat{\rho}_t=0|\rho_t=1] = \mathbb{P}\Big[ \hat{T}^{res}_{\chi^2,t} < \hat{T}_t\Big]\leq
\\&F^{\hat{\tau}^{cov}_t}_{\chi^2,p}\Big(\hat{T}_t\Big)
\Big(\Big[1-F^{gamma}_{r_{max},\epsilon_{cov},\Delta_l}(R_t\theta_l)\Big] + \Big[F^{ex}_{\Delta_l,\epsilon_{cov}}(\theta_l)\Big]^p\Big)
\end{aligned}
\end{equation}
\end{proof}
\vspace{-5mm}

\subsection{Proof of Theorem \ref{thm5}}\label{subsec:pf_thm5}
\begin{proof}
Based on the definition of $\epsilon$-DP, we know that the effective residual noise is given by $\hat{C}^{1/2}_we_w \sim N(0,\sigma^2\hat{S}_w)$ 
\begin{gather}\label{eq:thm51}
\Big|ln\Big[\frac{exp\big[-e^T_w\hat{C}^{-1}_we_w/(2\sigma^2)\big]}{exp\big[-(e_w+\Delta_r)^T\hat{C}^{-1}_w(e_w+\Delta_r)/(2\sigma^2)\big]} \Big]\Big| \leq \epsilon'
\end{gather}
Consolidating terms we obtain
\begin{gather}
\Big|\frac{\Delta_r(2.\mathbf{1}^T\hat{V}^T_w\hat{\Lambda}_w\hat{V}_we_w + \Delta_r\mathbf{1}^T\hat{S}_w\mathbf{1})}{2\sigma^2}\Big| \leq \epsilon'
\end{gather}
\begin{gather}\label{eq:thm52}
\Big|\mathbf{1}^T\hat{V}^T_w\hat{\Lambda}_w\hat{V}_we_w + \frac{\Delta_r\mathbf{1}^T\hat{S}_w\mathbf{1}}{2} \Big| \leq \frac{\sigma^2\epsilon'}{\Delta_r}
\end{gather}
Examining the LHS of Equation \eqref{eq:thm52}, we can see that
\begin{equation}\label{eq:thm52b}
\begin{aligned}
|\mathbf{1}^T\hat{V}^T_w\hat{\Lambda}_w\hat{V}_we_w| \leq ||\mathbf{1}^T\hat{V}^T_w\hat{\Lambda}_w||^2_2&||\hat{V}_we_w||^2_2 \leq\\ &(\mathbf{1}^T\hat{C}^{-1}_w\mathbf{1})^2||e_w||^2
\end{aligned}
\end{equation}
This implies that
\begin{equation}\label{eq:thm53a}
\begin{aligned}
\Big|\mathbf{1}^T\hat{V}^T_w\hat{\Lambda}_w\hat{V}_we_w + &\frac{\Delta_r\mathbf{1}^T\hat{S}_w\mathbf{1}}{2} \Big| \leq\\
& |\mathbf{1}^T\hat{V}^T_w\hat{\Lambda}_w\hat{V}_we_w| + \frac{\Delta_r\mathbf{1}^T\hat{C}^{-1}_w\mathbf{1}}{2}\\
\end{aligned}
\end{equation}
From Equation \eqref{eq:thm53a}, we obtain,
\begin{equation}\label{eq:thm53b}
\begin{aligned}
|\mathbf{1}^T\hat{V}^T_w\hat{\Lambda}_w\hat{V}_we_w| + & \frac{\Delta_r\mathbf{1}^T\hat{C}^{-1}_w\mathbf{1}}{2} \leq \\ 
& (\mathbf{1}^T\hat{C}^{-1}_w\mathbf{1})^2\Big(||e_w||^2+\frac{1}{2.\mathbf{1}^T\hat{C}^{-1}_w\mathbf{1}}\Big) 
\end{aligned}
\end{equation}

Characterizing the worst case (maximizing) $\epsilon'$ results in obtaining the upper bound
\begin{gather}
\frac{\sigma^2\epsilon'}{\Delta_r} \geq (\mathbf{1}^T\hat{C}^{-1}_w\mathbf{1})^2\Big(||e_w||^2+\frac{1}{2.\mathbf{1}^T\hat{C}^{-1}_w\mathbf{1}}\Big)\\
\implies \epsilon' \geq \frac{\Delta_r}{\sigma^2}(\mathbf{1}^T\hat{C}^{-1}_w\mathbf{1})^2\Big(||e_w||^2+\frac{1}{2.\mathbf{1}^T\hat{C}^{-1}_w\mathbf{1}}\Big) \label{eq:thm54}
\end{gather}
Using Lemma 2, we know that the probability that $||e_w||^2_2\geq\frac{\theta^2_r}{p}$ occurs is at most $1-(1-\gamma_r)^p$. Therefore, under conditions of Lemma 2, we can see that further maximizing $\epsilon'$ leads to a probabilistic lower bound of
\begin{equation}\label{eq:thm54a}
\begin{aligned}
\mathbb{P}\Big[ \epsilon' \geq \frac{\Delta_r}{\sigma^2}(\mathbf{1}^T\hat{C}^{-1}_w\mathbf{1})^2\Big(\frac{\theta^2_r}{p} +\frac{1}{2.\mathbf{1}^T\hat{C}^{-1}_w\mathbf{1}}\Big) & \Big] \leq\\
&1-(1-\gamma_r)^p
\end{aligned}
\end{equation}
\end{proof}

\subsection{Proof of Theorem \ref{thm6}}\label{subsec:pf_thm6}
\begin{proof}

\begin{table}[!htb]
\centering
\caption{Average Memory Usage (MB) and CPU Utilization (\%) for varying implementation modes}
\label{tab:system_design}
\resizebox{\columnwidth}{!}{%
\begin{tabular}{|l|l|c|c|c|c|}
\hline
\textbf{Dataset} & \textbf{Implementation} & \multicolumn{2}{c|}{\textbf{Average Memory (MB) (std dev.)}} & \multicolumn{2}{c|}{\textbf{CPU Utilization (\%) (std dev.)}} \\
\cline{3-6}
                 & \textbf{Mode} & Utility & Regulator & Utility & Regulator \\
\hline
\multirow{2}{*}{ORNL-PS}
    & Critical Region Verification & 1014.58 (224.85) & 128.96 (0.27) & 10.27 (12.34) & 10.55 (13.61) \\
    & P-Value Compliance & 1013.86 (223.18) & 128.29 (0.08) & 10.09 (12.07) & 9.91 (11.48) \\
\hline
\multirow{2}{*}{HAI}
    & Critical Region Verification & 1093.21 (221.24) & 128.80 (0.29) & 9.87 (11.15) & 9.93 (11.33) \\
    & P-Value Compliance & 1101.33 (271.88) & 128.02 (0.07) & 10.97 (13.38) & 10.47 (12.29) \\
\hline
\end{tabular}%
}
\end{table}

We consider the probability that $T^{cov}_w$ is obtained with a residual $r^1_w$ and DP-based eigenvalue matrix given by $\Lambda^1_w$. To establish differential privacy, we must now consider the probability of obtaining $T^{cov}_w$ when $r^2_w$ is realized and $\Lambda^2_w$ is the corresponding DP-based eigenvalue matrix. Using the definitions of DP, under conditions that $r^1_w$, $r^2_w$ and $\Lambda^1_w$, $\Lambda^2_w$ are adjacent pairs of realizations. Therefore without loss of generality we can state that:
\begin{gather}\label{eq:thm61}
ln\Big|\frac{exp(\frac{-r^TC^{-1}r}{2}).\mathbb{P}(\Lambda^1_w)}{exp(\frac{-(r+\Delta)^TC^{-1}(r+\Delta)}{2}).\mathbb{P}(\Lambda^2_w))}\Big| \leq \epsilon'
\end{gather}
Equation \eqref{eq:thm61} results in the following relationships
\begin{gather}
\Big|ln\Big(exp\Big[\frac{-2\Delta^TC^{-1}r + \Delta^TC^{-1}\Delta}{2\sigma^2}\Big].exp(\epsilon)\Big)\Big| \leq \epsilon'\\
\Big|\frac{-2\Delta^TC^{-1}r + \Delta^TC^{-1}\Delta}{2\sigma^2} +\epsilon \Big| \leq \epsilon' \label{eq:thm62}
\end{gather}
We can now use $\epsilon'$ to bound the RHS such that the following holds
\begin{gather}
||r|| \leq \frac{\sigma^2(\epsilon'-\epsilon)}{||\Delta^TC^{-1}||}-\frac{\Delta^TC^{-1}\Delta}{2||\Delta^TC^{-1}||}
\end{gather}
Since $r\sim N(0,C)$, we can state that $\Delta^TC^{-1}r \sim N(0,\Delta^TC^{-1}\Delta)$.  As a consequence we obtain the following relationship where $\Phi$ is the CDF for $N(0,\Delta^TC^{-1}\Delta)$
\begin{equation}\label{eq:thm63}
\begin{aligned}
\mathbb{P}\Big[||r|| \leq &\frac{\sigma^2(\epsilon'-\epsilon)}{||\Delta^TC^{-1}||}-\frac{\Delta^TC^{-1}\Delta}{2||\Delta^TC^{-1}||} \Big] =\\ 
&\Phi\Big(\frac{\sigma^2(\epsilon'-\epsilon)}{||\Delta^TC^{-1}||}
- \frac{\Delta^TC^{-1}\Delta}{2||\Delta^TC^{-1}||} \Big)\\
- & \Phi\Big(-\frac{\sigma^2(\epsilon'-\epsilon)}{||\Delta^TC^{-1}||}+\frac{\Delta^TC^{-1}\Delta}{2||\Delta^TC^{-1}||}\Big)
\end{aligned}
\end{equation}
To ensure that privacy loss associated with disclosure of $\hat{T}^{cov}_w$ is bounded by $\epsilon'$ we would need a $\delta'$ such that 
\begin{equation}\label{eq:thm64}
\begin{aligned}
\delta'\leq \Phi\Big(\frac{\sigma^2(\epsilon'-\epsilon)}{||\Delta^TC^{-1}||}-&\frac{\Delta^TC^{-1}\Delta}{2||\Delta^TC^{-1}||} \Big) \\- & \Phi\Big(-\frac{\sigma^2(\epsilon'-\epsilon)}{||\Delta^TC^{-1}||}+\frac{\Delta^TC^{-1}\Delta}{2||\Delta^TC^{-1}||}\Big)
\end{aligned}
\end{equation}
\end{proof}

\section{Supplementary Information on Experiments}

\noindent
\subsection{Dataset and Detection Models}\label{subsec:ddm}
For our experiments, we primarily leverage the HAI Security dataset \cite{hai}, as well as the ORNL power system (ORNL-PS) attack dataset \cite{pan2015classification,pan2015developing}. For both datasets, we trained an NLKF model as described in Section \ref{subsec:nlkfe} using the corresponding state and sensor variables pertaining to each dataset. In order to yield a well-formed non-linear, extended Kalman Filter model for the HAI dataset, we utilized the multi-level LSTM framework provided in \cite{coskun2017long}. Our experimental strategy revolves around evaluating the variations pertaining to DP failure probabilities $\gamma_{cov},\gamma_r$, the privacy budgets $\epsilon_{cov},\epsilon_r$ respectively. 

\noindent
\subsection{System Implementation Details}\label{subsec:sys_det}
All experiments were carried out on a virtual machine (VM) running Ubuntu 24.04 with 100GB of RAM and 16 vCPUs using \texttt{Python 3.11} with the detection model inference using \texttt{PyTorch 2.7.1}. For evaluating diverse aspects of our proposed framework, we utilized a native as well as a container based execution environment. The native setup was primarily used for evaluating the performance of our framework under various scenarios of differential privacy. The container based execution environment was used to evaluate the distinct implementation modes pertaining to Critical Region Verification (CRV) and the P-Value Compliance (PVC). Specifically, we generated container images representative of the regulator and the utility that replays snippets of the HAI and ORNL-PS datasets under scenarios of attack. In order to evaluate system performance of CRV and PVC mode, we created container images representative of the regulator and the utility that replays attack scenarios of HAI and ORNL-PS datasets. The regulator container service hosts a REST API developed using \texttt{Flask} that receives DP-driven disclosures from the corresponding utility container service for each implementation mode and executes the corresponding compliance steps as provided in Algorithms \ref{alg:dpalg1_rl} and \ref{alg:dpalg4_rl}. Consequently, we measure the system performance in terms of the CPU utilization and memory consumption of both utility and regulator services under both implementation modes. Docker based quick start scripts and the associated code have been provided as part of the accompanying artifacts to our paper.


\section{Supplementary Results}

\subsection{Analyzing System Performance}\label{subsec:sys_imp}
In Table \ref{tab:system_design}, we provide a comparative analysis of system performance of the CRV and PVC implementation modes with respect to HAI and ORNL-PS datasets.  Average CPU utilization consistently stays under 11\% for all scenarios with standard deviation ranging from 11.15\% to 13.6\%. Similarly, memory usage for utility service ranges between 1014 MB and 1101 MB, while remaining very close to 128 MB for the regulator service. From Table \ref{tab:system_design}, we also see that the standard deviation values of CRV memory usage is relatively higher for both datasets. This rise can be explained on the basis of the need to factorize the DP-driven covariance matrix by the regulator in the CRV mode as opposed to a simple compliance check needed in the PVC mode. Overall, Table \ref{tab:system_design} demonstrates that the system performance in terms of both average memory consumption as well as CPU utilization remain consistent and stable across both datasets and implementation modes. The codebase is publicly available at \url{https://github.com/disys-lab/precise.git}.

\end{document}